\documentclass[authoryear,preprint,review,12pt]{elsarticle}

\usepackage{amssymb}
\usepackage{amsmath}

\journal{Electronic Commerce Research and Applications}

\begin{document}

\begin{frontmatter}



\title{Orchestrating Rewards in the Era of Intelligence-Driven Commerce} 


\author[aff1,aff2]{Paul Osemudiame Oamen} 
            
\author[aff1]{Robert Wesley} 
	
\author[aff3]{Pius Onobhayedo} 

\affiliation[aff1]{organization={My AI Inc.},
	addressline={8 The Green. Suite}, 
	city={Dover},
	postcode={11901}, 
	state={DE},
	country={USA}}
	
\affiliation[aff2]{organization={School of Natural and Computing Sciences, University of Aberdeen},
	city={Aberdeen},
	postcode={AB24 3FX}, 
	state={Scotland},
	country={UK}}
	
\affiliation[aff3]{organization={Marshall School of Business, University of Southern California},
	city={Los Angeles},
	postcode={90007}, 
	state={CA},
	country={USA}}

\begin{abstract}
Despite their evolution from early copper-token schemes to sophisticated digital solutions, loyalty programs remain predominantly closed ecosystems, with brands retaining full control over all components. Coalition loyalty programs emerged to enable cross-brand interoperability, but approximately 60\% fail within 10 years in spite of theoretical advantages rooted in network economics. This paper demonstrates that coalition failures stem from fundamental architectural limitations in centralized operator models rather than operational deficiencies, and argues further that neither closed nor coalition systems can scale in intelligence-driven paradigms where AI agents mediate commerce and demand trustless, protocol-based coordination that existing architectures cannot provide.

We propose a hybrid framework where brands maintain sovereign control over their programs while enabling cross-brand interoperability through trustless exchange mechanisms. Our framework preserves closed system advantages while enabling open system benefits without the structural problems that doom traditional coalitions. We derive a mathematical pricing model accounting for empirically-validated market factors while enabling fair value exchange across interoperable reward systems.
\end{abstract}

\begin{graphicalabstract}
\end{graphicalabstract}



\end{frontmatter}



\section{Introduction}
\label{introduction}
Loyalty rewards have proven remarkably enduring as commercial instruments, adapting their form while preserving a fundamental principle that strategic incentives shape consumer behavior and drive business outcomes \citep{kumar2004building}. From the copper tokens American merchants issued in the 1890s to today's sophisticated digital ecosystems embedded in mobile applications, what has evolved least significantly is the degree of openness in their exchange and redemption.

The modern loyalty program emerged in 1981 when American Airlines launched AAdvantage, recognizing that business travelers making flight decisions for their employers could be personally incentivized through individual rewards \citep{bainbridge1997cashing}. Within three years, competing airlines had established similar programs, understanding that absence from the loyalty space created competitive disadvantage \citep{gilbert1996relationship}. Digital platforms and mobile technology accelerated this evolution, dramatically reducing operational friction and enabling real-time engagement at the point of purchase.

Throughout this evolution, however, loyalty programs remained predominantly closed systems where individual brands control all aspects of reward issuance and redemption within their own ecosystems. Industry research reveals that consumers belong to an average of 14.8 loyalty programs but actively participate in only 6.7, suggesting that fragmentation across isolated programs leaves substantial value unrealized \citep{bond2020loyalty}. This observation naturally raised the question of whether rewards could deliver greater value through interoperability across multiple businesses, representing the open systems vision.

Coalition loyalty programs emerged as the dominant attempt to implement this vision, with multiple brands partnering through central operators who manage shared currencies that customers earn and redeem across partner networks. The theoretical advantages appeared compelling, rooted in network economics, economies of scale, and enhanced customer value propositions. The empirical reality, however, proved starkly different. Industry analysis reveals that approximately 60\% of coalition programs fail within 10 years of launch, with high-profile examples including American Express's Plenti, which shut down after three years despite over \$100 million in investment \citep{bond2020loyalty,shoulberg2018plenti}.

This widespread failure pattern raises a fundamental question that motivates our research. Do coalitions face merely operational challenges that better management could resolve, or do they suffer from structural problems inherent in their architectural design? Our systematic analysis reveals that coalition failures stem from fundamental architectural limitations in centralized operator models rather than operational deficiencies. These structural problems grow increasingly critical as commerce transforms, with artificial intelligence systems now mediating relationships between consumers and businesses through shopping assistants, browser extensions, and autonomous purchasing agents \citep{russell2020artificial}. When intelligent agents require standardized, machine-readable interfaces for effective operation, the architectural limitations of traditional coalition models become absolute barriers rather than mere inefficiencies.

This paper makes three primary contributions to loyalty program literature and practice. First, we provide systematic analysis demonstrating that coalition failures stem from architectural limitations rather than operational deficiencies, identifying the centralized operator model as the root cause. Second, we propose a hybrid framework where brands maintain sovereign control over their programs while enabling cross-brand interoperability through trustless exchange mechanisms implemented via smart contracts and automated market makers. Third, we derive a mathematical pricing model that accounts for empirically-validated market factors while enabling fair value exchange in decentralized reward ecosystems.

Our framework preserves the advantages of closed systems while enabling the benefits of open systems without the structural problems that doom traditional coalitions. By eliminating central operators in favor of protocol-based coordination, we address what we identify as the root cause of coalition failures, namely intermediary rent-seeking and misaligned incentives that cannot be resolved through operational improvements alone.

\section{Background and Related Work}
\label{sec:background}

Understanding why coalition loyalty programs fail requires grounding in both the theoretical foundations that explain what makes loyalty programs effective and the empirical realities of how coalitions have operated in practice. This section synthesizes relevant literature across loyalty program theory, the evolution of closed and open systems, documented coalition failures, intelligence-driven commerce requirements, and the properties of trustless systems that enable alternative architectural approaches.

\subsection{Loyalty Program Theory and Effectiveness}
\label{subsec:loyalty_theory}

The academic study of loyalty programs has produced substantial insights into why and how reward mechanisms influence consumer behavior. At their foundation, loyalty programs operate through principles of behavioral conditioning, where repeated rewards for desired behaviors create associations that increase the probability of future similar behaviors \citep{skinner1953science}. This psychological foundation manifests in commercial contexts as increased purchase frequency, higher average transaction values, and longer customer relationships \citep{kumar2004building}.

The effectiveness of loyalty programs stems from multiple psychological mechanisms working in concert. Goal-gradient effects demonstrate that customers accelerate their purchasing as they approach reward thresholds, a phenomenon extensively documented in both laboratory and field settings \citep{kivetz2006goal}. The endowment effect creates psychological ownership of accumulated points, making customers reluctant to abandon programs even when rational economic analysis might suggest doing so \citep{kahneman1991anomalies}. Sunk cost fallacy further reinforces continued participation as customers feel they have already invested effort into accumulating rewards \citep{thaler1980toward}. These psychological mechanisms combine to create switching costs that extend beyond purely economic considerations into emotional and cognitive domains.

Research on program design has identified several factors that significantly influence effectiveness. Personalization emerges consistently as a critical driver of engagement, with programs that tailor offers to individual preferences and behaviors demonstrating substantially higher participation rates and customer satisfaction than generic programs \citep{kumar2016creating}. The structure of reward schedules matters considerably, with research showing that transparent, easily understood earning and redemption mechanisms outperform complex tiered structures that obscure true value \citep{nunes2006your}. Redemption flexibility proves essential, as customers value the ability to use rewards in ways aligned with their actual preferences rather than being constrained to limited redemption catalogs \citep{dorotic2012loyalty}.

The relationship between loyalty programs and actual behavioral loyalty, however, proves more nuanced than simple reward-response models would suggest. While programs demonstrably increase repeat purchasing among participants, questions persist about causality. Do programs create loyalty or merely identify and reward customers who would have been loyal regardless? Research addressing this self-selection problem through careful methodological controls finds that programs do generate incremental behavioral change beyond what would occur absent the program, though the magnitude varies considerably across contexts \citep{leenheer2007do}. Critically, single-brand programs operated by large retailers with frequent purchase occasions tend to show stronger effects than multi-brand coalitions, suggesting that program structure influences effectiveness substantially.

\subsection{The Evolution of Closed and Open Systems}
\label{subsec:closed_open_evolution}

The evolution of loyalty programs has produced two dominant paradigms representing fundamentally different approaches to reward system design. Understanding these alternatives provides essential context for analyzing coalition failures and proposing hybrid solutions.

\subsubsection{Closed Systems}

Closed Systems represent the traditional single-brand model where one company controls all aspects of reward issuance, management, and redemption within its own ecosystem. These systems exhibit several key characteristics, including complete brand autonomy over earning rates, redemption values, and program rules. Brands maintain direct customer relationships and own all data generated through program interactions. This enables focused personalization based on detailed behavioral data from repeated transactions. Additionally, closed systems preserve clear brand identity where customers perceive rewards as coming from and reinforcing relationships with specific companies \citep{keller1993conceptualizing}.

The advantages of closed systems are substantial and well-documented. Brands maintain full control over program economics, adjusting reward generosity based on contribution margins and competitive positioning. They capture all customer data, enabling sophisticated personalization and strategic insights that drive program effectiveness \citep{kumar2016creating}. The direct brand-customer relationship creates emotional bonds that strengthen over time, building the psychological attachment that drives repeat purchasing behavior. Program rules can be optimized for specific business models without compromise for coalition partners.

\begin{figure}[t]
	\centering
	\includegraphics[width=0.5\textwidth]{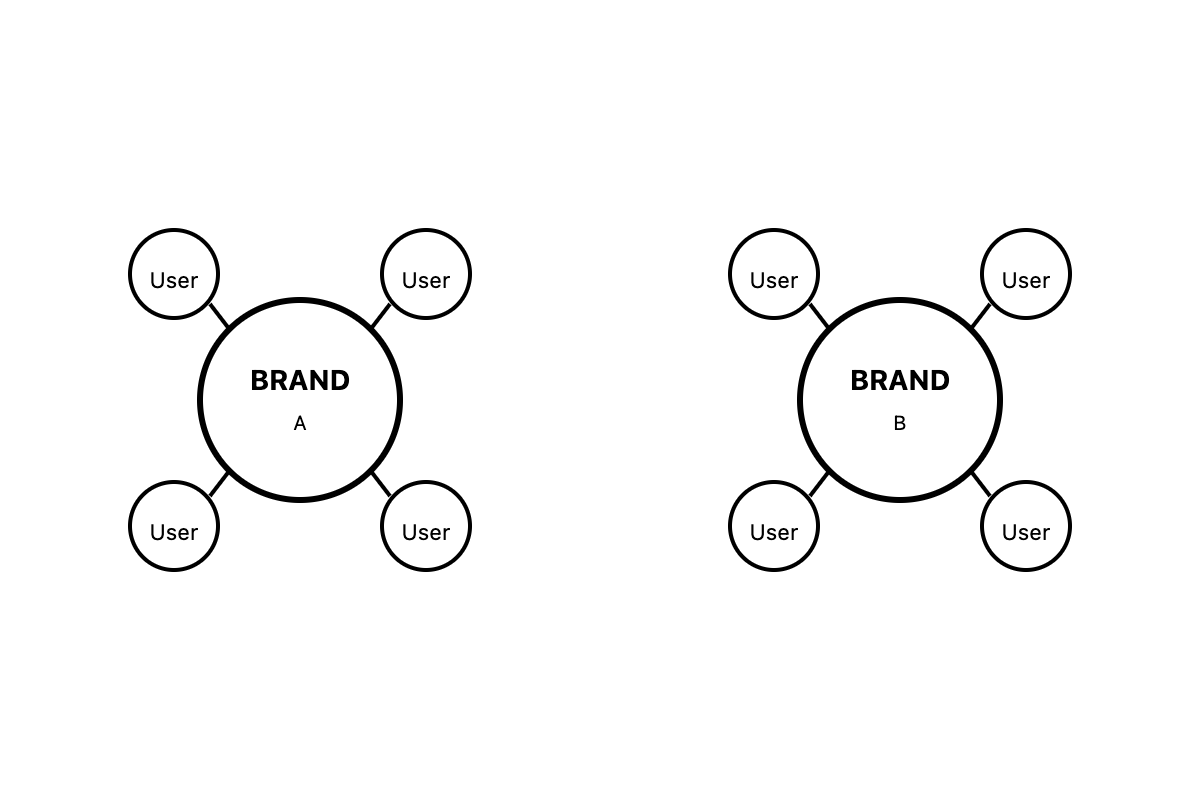}
	\caption{Closed Loyalty System Model. Two brands operate independent loyalty programs with complete ecosystem isolation. Lines indicate bilateral transaction-reward relationships within each siloed system. User profiles are brand-specific with no cross-brand recognition; even if the same individual participates in both programs, each brand maintains separate accounts, balances, and behavioral data with no interoperability or shared value proposition.}
	\label{fig:closed_system}
\end{figure}

However, closed systems face inherent limitations that restrict their value proposition. Customers experience fragmentation, managing multiple programs with separate accounts, balances, and rules, creating cognitive burden and reducing engagement \citep{nunes2006your}. Limited network effects emerge because value remains constrained to single-brand touchpoints, preventing the accumulation synergies that multi-brand systems could theoretically provide. Smaller businesses struggle with economies of scale, finding the infrastructure investments and operational costs prohibitive relative to their customer base and transaction volumes. Breakage (unredeemed points) may be high when customers lack sufficient engagement with single brands to accumulate meaningful balances, representing dead-weight loss in the system. Most fundamentally, closed systems misapprehend the reality that users are inherently social beings. The same individual who flies an airline also dines at restaurants, shops at malls, stays in hotels, and attends sporting events. While such systems may suffice in today's landscape, they face inevitable obsolescence as thousands of businesses compete for increasingly scarce user attention. In this fragmented environment, requiring users to independently manage rewards across disparate platforms becomes untenable. Moreover, in an emerging AI-driven agentic paradigm capable of computing complex user preferences instantaneously, closed systems dramatically underutilize optimization capabilities. Strategies that optimize across composite user goals, treating users as the social entities they are rather than isolated economic actors, will demonstrably outperform siloed approaches that fail to reflect the interconnected nature of human behavior.

\subsubsection{Open Systems (Coalitions)}

Open Systems (coalitions) emerged to address these limitations through multi-brand partnerships coordinated by central operators \citep{sharp1997loyalty}. These systems promise several theoretical advantages rooted in network economics. Enhanced customer value should emerge through the ability to earn and redeem across multiple brands, accelerating accumulation and increasing utility through broader redemption options. Network effects suggest that each additional partner increases value for all participants, creating positive feedback loops that strengthen the ecosystem \citep{katz1985network}. Cost sharing enables brands to split infrastructure, marketing, and operational expenses, making loyalty programs economically viable for smaller participants. Cross-brand acquisition promises that partners gain access to customers from other coalition members, expanding their addressable market.

The coalition model involves three primary actors whose interactions define system dynamics \citep{dorotic2012loyalty}. The central operator manages technology infrastructure, customer databases, point accounting systems, and partner relationships, typically operating as a for-profit entity charging fees for these services. Member brands agree to award coalition points for purchases and accept coalition points for redemption through contractual arrangements that specify earning rates, redemption values, fees, and data sharing provisions. Customers interact primarily with the coalition brand rather than individual partners, earning a generic currency usable across the partner network. 

From the customer perspective, coalitions theoretically offer simplified participation through a single account and currency, increased redemption flexibility across multiple brands and categories, faster accumulation when points from various purchases combine, and greater redemption value through more options and better matching to preferences.

\begin{figure}[t]
	\centering
	\includegraphics[width=0.5\textwidth]{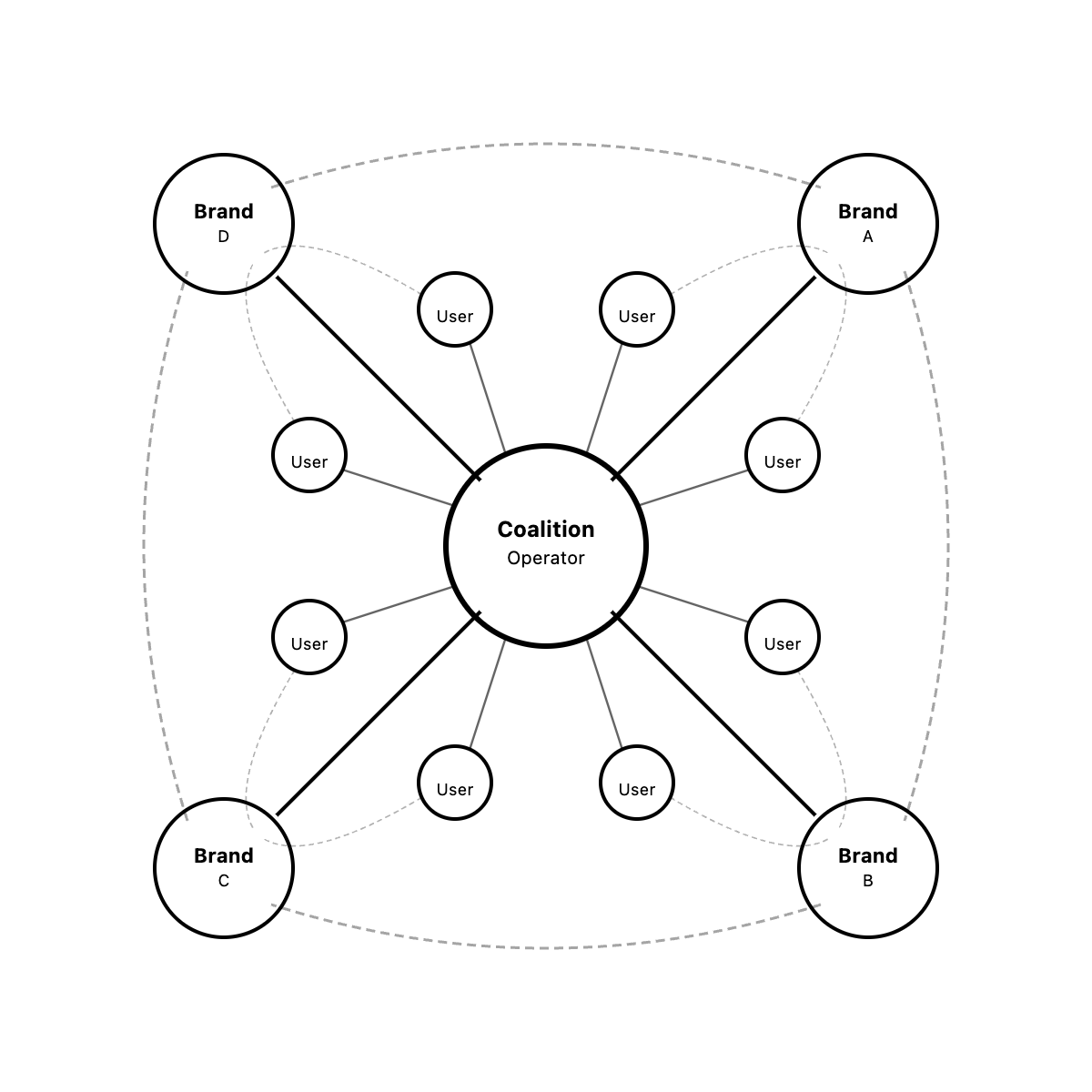}
	\caption{Coalition Loyalty System Model. Multiple brands participate through contractual relationships with a central operator who manages infrastructure, point accounting, and unified customer databases. Users maintain consolidated profiles (solid lines to operator) recognized across all coalition partners. Faint dashed connections to individual brands represent likely entry points through which users initially accessed the coalition ecosystem. Network effects between brands (perimeter connections) demonstrate the interdependent value creation mechanism. Unlike closed systems, users access multiple brands through a single identity while brands share infrastructure costs and customer data.}
	\label{fig:coalition_system}
\end{figure}


\subsection{Coalition Program Failures: Empirical Evidence and Structural Analysis}
\label{subsec:coalition_failures}

While the coalition model presents compelling theoretical advantages, empirical evidence reveals consistent patterns of underperformance and failure that suggest architectural rather than operational problems. We examine three representative historical failures before presenting systematic analysis of the underlying structural challenges.

\subsubsection{Empirical Evidence from Coalition Failures}

Air Miles Canada, one of the longest-running coalition programs, grew to over 10 million member households and more than 100 brand partners over two decades, achieving scale that should have generated the network effects and economies central to coalition value propositions. However, the program faced persistent profitability challenges that led to a critical 2016 decision where announcing that points would begin expiring triggered massive customer backlash, government scrutiny, and intense negative media coverage \citep{cbc2016airmiles,cbc2016airmiles1}. The coalition operator was ultimately forced to reverse the policy, but the incident severely damaged program trust and exposed fundamental tensions between coalition economics and customer satisfaction. While Air Miles survived this crisis, the experience revealed that even large-scale, established coalitions struggle with economic sustainability.

Nectar in the United Kingdom illustrates the challenging economics underlying large coalition loyalty programs. Launched in 2002 with anchor partners Sainsbury’s, BP, Barclaycard, and Debenhams, Nectar rapidly grew to nearly 13 million members within its first year and is now described by Nectar360 as the UK’s largest coalition loyalty program \citep{warc2003uk, nectar3602024nectar}. In the years leading up to 2018, the takeover of partner Homebase by Bunnings and the exit of British Gas reduced the program’s ability to add new large non-competitive partners, and Aimia, Nectar’s operator at the time, concluded that retaining ownership offered limited opportunities to create value \citep{aimia2018aimiasells}.

In February 2018, Aimia sold Nectar and related UK assets to Sainsbury’s for gross consideration of approximately Can\$105 million (£60 million), while simultaneously transferring approximately Can\$183 million (£105 million) of cash to cover redemption liabilities and approximately Can\$96 million (£55 million) of net working capital. After giving effect to the transaction, Aimia reported that its net cash and liquidity position declined by approximately Can\$174 million \citep{aimia2018aimiasells}. Notably, the divested business had generated Can\$57.1 million in Adjusted EBITDA and Can\$49.9
million in Free Cash Flow during 2017 \citep{aimia2018aimiareports}. The deal structure suggests that covering accumulated redemption obligations required cash transfers substantially exceeding the headline sale price. Sainsbury’s, already Nectar’s largest issuance and redemption partner across many key retail categories, was better positioned than independent operators to manage these liabilities on its own balance sheet. This transaction illustrates that even successful coalition programs can accumulate redemption obligations requiring substantial balance-sheet capacity, placing the operation
of large-scale coalitions beyond the reach of firms without deep pockets.

Plenti in the United States represents perhaps the most dramatic coalition failure. Backed by American Express and featuring high-profile partners including AT\&T, Exxon, Macy's, and Rite Aid, Plenti launched in 2015 with over \$100 million in marketing investment and ambitious projections to establish the first major multi-brand loyalty program in the U.S. market \citep{shoulberg2018plenti}. However, within only three years, American Express announced the program's shutdown, citing inability to achieve projected growth and engagement targets. The rapid failure despite enormous resources, sophisticated operator expertise, and prominent partners demonstrated conclusively that coalition challenges extend beyond operational execution to fundamental structural issues that cannot be resolved through better management or increased investment.

\subsubsection{Nine Fundamental Structural Challenges}

Systematic analysis of these and other coalition failures reveals nine interconnected challenges that explain why centralized operator models cannot deliver sustainable open reward systems. These are not independent operational problems but manifestations of deeper architectural limitations.

\textbf{Brand Identity Erosion.} Coalition programs require participating brands to award generic coalition currency rather than brand-specific rewards. When customers shop at Starbucks within a coalition, they earn "Coalition Points" rather than "Starbucks Stars." This homogenization serves the functional purpose of enabling interoperability, but simultaneously strips away the brand-specific identity that makes rewards psychologically valuable and emotionally resonant. Brand identity represents a carefully cultivated set of associations, values, and emotional connections that differentiate one company from competitors \citep{keller1993conceptualizing}. When rewards become generic coalition points, the brand disappears from the reward experience. Customers perceive themselves as earning value from the coalition operator rather than from the specific brand where the transaction occurred. Research comparing coalition member satisfaction with single-brand program satisfaction consistently finds that coalition participants report lower emotional attachment to individual brands and higher price sensitivity \citep{dorotic2014reward}, demonstrating that identity erosion produces measurable degradation in program effectiveness.

\textbf{Partner Cannibalization.} When multiple brands within a coalition compete for overlapping customer segments, rewards earned at one partner are often redeemed at competitors, creating value transfers between rivals rather than genuine network benefits. While coalitions carefully avoid placing direct head-to-head competitors in identical categories, substitution effects prove broader than simple product category definitions. Restaurants within a coalition compete for dining occasions, retail brands compete for discretionary spending, and entertainment venues compete for leisure time and budgets. Research examining coalition effects on purchase behavior finds that transaction increases at one partner correlate with decreases at other partners within the same coalition \citep{meyerwaarden2009grocery}, suggesting that much coalition activity represents wallet share redistribution rather than demand expansion. Coalition operators benefit from total transaction volume regardless of distribution among partners, creating misaligned incentives where operators have little motivation to prevent cannibalization while individual brands suffer from it directly.

\textbf{Data Ownership Asymmetry.} Coalition structures require customers to register with the coalition rather than with individual brands. The coalition operator creates customer accounts, maintains the master customer database, and tracks all transactions across the partner network. Brands receive aggregate reports and limited segmented data but typically cannot access individual customer transaction histories or personally identifiable information. This asymmetry creates multiple problems. Brands cannot effectively personalize their offerings because they lack detailed behavioral data. They cannot independently verify coalition reports about program performance, creating information disadvantage in operator relationships. They cannot build proprietary customer insights that might inform product development or strategic positioning. Most critically, if brands exit coalitions, they lose all customer relationships and data accumulated during participation. The data hostage problem creates significant exit barriers, enabling coalition operators to maintain participation from dissatisfied brands who have made sunk investments in customer relationships they cannot take with them.


\textbf{Reputation Contamination.} Coalition programs create shared brand identities where all partners associate themselves with the coalition promise. This shared branding extends the coalition's reputation to all members and vice versa, creating contagion vectors when individual partners experience service failures. Customers often cannot clearly distinguish between failures attributable to specific brands versus coalition operations, leading to attribution ambiguity where negative experiences at one partner spread reputational damage across the entire network. Empirical evidence documents significant contamination effects where customer satisfaction scores decline across coalitions when major partners face scandals or service failures, including for partners uninvolved in the specific incident \citep{schumann2014targeted}. Premium brands sharing coalition membership with discount or low-quality partners risk reputational contamination, representing one reason such brands typically avoid coalition participation.

\textbf{Complexity and User Experience.} Coalition programs are inherently complex coordination systems. Customers must understand which brands participate, how points are earned across different categories, what exchange rates apply between partners, where and how redemption occurs, what restrictions and blackout periods apply, and how value compares across different usage options. This complexity creates friction that reduces engagement and limits program effectiveness. Research consistently finds that 73\% of consumers find loyalty programs too complicated, 54\% remain unaware of which programs they belong to, and 45\% of coalition members remain inactive \citep{bond2020loyalty}. Coalition operators face impossible tradeoffs between comprehensiveness and simplicity. Broad partner networks increase value but multiply complexity, partner-specific rules accommodate brand requirements but confuse customers, and promotional campaigns drive engagement but add temporal complexity. Attempts to simplify often degrade into oversimplification that fails to communicate actual program mechanics, leading to customer disappointment when expected value does not materialize or unexpected restrictions apply.

\textbf{Misaligned Economic Incentives.} Coalition programs involve three parties with fundamentally different and conflicting objectives. Coalition operators must extract sufficient value from the ecosystem to cover substantial operational costs including technology infrastructure, marketing, customer service, and corporate overhead, typically through fees ranging from 10\% to 30\% of transaction value \citep{capizzi2005loyalty}. Brands seek customer acquisition and retention at acceptable cost, questioning whether coalition participation generates genuine incremental value or merely redistributes existing customer spending while enriching the operator. Customers seek to maximize the value they extract from accumulated points, creating adverse selection where the most engaged users extract maximum value relative to spending, degrading program economics. The fundamental problem is that coalition operators thrive by standing between brands and customers, extracting value from both parties. The more successful the operator becomes at monetizing its intermediary position, the less value remains for brands and customers, creating antagonistic rather than cooperative relationships that cannot be resolved through operational adjustments.

\textbf{Lost Personalization.} Effective loyalty programs tailor offers and communications to individual customer preferences, purchase histories, and predicted needs, with personalization emerging as a critical driver of engagement and program effectiveness \citep{kumar2016creating}. Coalition programs systematically underperform on personalization because the required data flows through operator infrastructure, gets aggregated across multiple brands, and returns to brands as delayed, aggregated reports if at all. Coalition operators face conflicting pressures. Brands want maximal data access to enable personalization, but operators' business models often depend on monetizing data, creating incentives to restrict brand access. The practical result is that brands within coalitions receive insufficient data to personalize effectively, knowing perhaps that a customer earned points recently but not what specifically was purchased, or seeing category-level summaries but not item-level detail. The inability to personalize manifests in customer experience as generic rather than tailored communications, with coalition promotions tending toward broad category multipliers rather than product-specific offers matched to individual preferences.

\textbf{Value Dilution and Exclusivity Loss.} Premium brands build market positions through cultivated exclusivity, where limited availability creates psychological value beyond functional utility \citep{keller1993conceptualizing}. When coalition programs make rewards broadly available across mass-market partners, they threaten the exclusivity that premium brands have carefully constructed. Premium brands joining coalitions that include mass-market partners risk associating their names with lower-tier offerings in ways that erode positioning. Coalition structures struggle to accommodate varying exclusivity strategies, with operators seeking to maximize network size through broad participation while premium brands prefer selective partnerships with other premium brands. The resulting tension typically leads to either exclusion of premium brands (reducing coalition appeal to affluent customers) or inclusion on terms that compromise exclusivity (reducing coalition appeal to premium brands).

\textbf{Unsustainable Economic Model.} Coalition programs face persistently challenging economics that have proven fatal for most implementations. The centralized operator model requires substantial upfront investment ranging from \$50 million to \$500 million depending on market size and ambitions, covering technology infrastructure, customer acquisition marketing, partner onboarding, and brand building. Ongoing operational costs compound the sustainability challenge, with technology maintenance, staffing, marketing, and customer service operations creating annual cash burn rates of \$50-100 million for meaningful coalitions \citep{leenheer2007do}. Revenue generation proves slower than cost accumulation, with transaction fees providing primary income but requiring years of growth to reach break-even volume. Industry analysis reveals sobering statistics where approximately 60\% of coalitions fail within 10 years, never achieving sustainable operations, while those that survive to profitability typically require 5-7 years and \$300-700 million in cumulative investment \citep{bond2020loyalty}. The fundamental problem is that centralized operator models have limited economies of scale, with costs growing nearly proportionally with operational complexity.

\textbf{Structural Root Cause.} These nine challenges are not independent problems but interconnected manifestations of a deeper architectural limitation, specifically the centralized operator model that positions an intermediary between brands and customers. This architectural choice generates value extraction opportunities for operators (who capture fees, control data, and exercise discretionary authority) but creates systematic problems for all other participants. Brands lose identity, data, and control while paying fees. Customers face complexity, reduced personalization, and value dilution. Partners compete within the coalition while subsidizing the operator. The fundamental misalignment cannot be resolved through operational improvements because it stems from the architecture itself, suggesting that alternative architectural approaches are necessary for sustainable open reward systems.


\subsection{Intelligence-Driven Commerce and the Imperative for Open Rewards}
\label{subsec:intelligence_driven}

The case for resolving coalition structural problems becomes urgent as commerce undergoes fundamental transformation. Artificial intelligence systems now mediate relationships between consumers and businesses, augmenting human decision-making through tools that execute preferences, optimize choices, and manage transactions at speeds and scales impractical for manual achievement \citep{russell2020artificial}.

Shopping assistants such as Amazon's Alexa and Google Assistant receive verbal instructions to purchase products, compare prices across retailers, and complete transactions autonomously \citep{davenport2020artificial}. Browser extensions automatically apply discount codes and track price histories to identify optimal purchase timing. Robo-advisors make portfolio allocation decisions based on risk preferences and market conditions. Smart home systems autonomously reorder consumable supplies when inventories run low, managing household logistics without continuous human oversight \citep{maass2017cognitive}. These intelligent agents function as extensions of human intention, programmed to pursue objectives defined by their users.

The implications for commercial systems are profound. When humans evaluate options directly, certain inefficiencies prove tolerable. Limited information processing capacity means consumers typically compare only a small number of alternatives, while search costs naturally constrain optimization efforts \citep{simon1955behavioral, stigler1961economics}. Intelligent agents fundamentally alter these dynamics. Systems that query hundreds of data sources simultaneously, process complex optimization problems in milliseconds, and maintain perfect recall of preferences and transaction histories transform what becomes practically achievable \citep{brynjolfsson2014second}.

This transformation has crucial implications for reward systems. Features providing value but proving impractical for humans to exploit fully become genuinely useful when agents leverage them effortlessly. Where a consumer might check three stores before purchasing, an agent evaluates every market option, incorporating rewards availability across accessible programs to identify globally optimal choices. The ability to use rewards across dozens of partner businesses, optimize redemption timing based on fluctuating exchange rates, and identify optimal combinations of rewards and pricing shifts from theoretical advantage to practical capability exploitable at scale. Conversely, characteristics that humans tolerate become intolerable barriers when agents require standardized, machine-readable interfaces. Opaque pricing, manual processes, and complex rules that merely inconvenience humans prevent algorithmic optimization entirely, potentially excluding businesses from agent-mediated commerce. Open reward systems, with their cross-platform optionality and network effects, align naturally with this intelligence-driven paradigm.

\subsection{Trustless Perpetual Systems: Architectural Foundations for Alternative Coordination}
\label{subsec:trustless_systems}

Understanding potential alternatives to traditional coalition structures requires examining trustless systems and their distinctive properties for coordinating value exchange without intermediaries. These architectural patterns emerged from distributed systems research and have been applied successfully in domains requiring coordination without central authority.

\textbf{Distributed Ledger Technology} provides infrastructure where no single entity controls operations. A distributed ledger maintains shared transaction records replicated across multiple independent nodes, with cryptographic techniques ensuring records cannot be altered retroactively and all participants observe identical history \citep{nakamoto2008bitcoin}. This architecture eliminates single points of failure, provides transparency through independently verifiable transaction histories, and creates immutability as altering records requires corrupting a majority of nodes simultaneously, which becomes computationally or economically infeasible in properly designed systems \citep{swan2015blockchain}.

\textbf{Smart Contracts} extend distributed ledgers by enabling self-executing code that automatically enforces agreed-upon rules \citep{szabo1997formalizing}. Rather than relying on institutions to execute contracts, smart contracts embed logic directly in blockchain infrastructure where it executes automatically when conditions are met. This eliminates intermediaries for many functions, reduces transaction costs, and guarantees execution as properly designed contracts cannot be prevented from running by any single party.

\textbf{Automated Market Makers} (AMMs) represent a specific smart contract application particularly relevant to reward systems. AMMs use algorithmic formulas to determine exchange rates between assets based on supply and demand reflected in liquidity pools \citep{adams2020uniswap}. Participants provide liquidity by depositing assets, earning fees in exchange, while others execute trades at algorithmically determined rates. This mechanism enables decentralized exchange without operators setting prices, matching counterparties, or maintaining order books.

These technical primitives enable coordination through protocols rather than operators. Where traditional coalitions centralize functions in fee-extracting intermediaries exercising discretionary control, trustless systems coordinate behavior through permissionless, transparent, automated mechanisms embedded in smart contracts that no party controls unilaterally. As autonomous intelligence oracles using AI and machine learning mature, they will further enhance capabilities through sophisticated real-time data feeds. The question becomes whether these architectural alternatives can address the nine structural problems fatal to traditional coalition programs while preserving coordination benefits that make open reward systems theoretically attractive. We turn to this question in the following sections, proposing a hybrid framework that combines the sovereignty of closed systems with the interoperability of open systems, implemented through trustless coordination mechanisms rather than centralized operators.

\section{The Hybrid Model: Theoretical Foundations}
\label{sec:theoretical_foundations}

\subsection{Core Proposition: The Closed-Open Architecture}
\label{subsec:core_proposition}

Our analysis in Section~\ref{sec:background} revealed that coalition failures stem from a fundamental architectural flaw where centralized operators position themselves between brands and customers. This suggests that sustainable open reward systems require fundamentally different architecture. We propose a hybrid framework that preserves brand sovereignty while enabling cross-brand interoperability, combining what we term ``closed autonomy'' with ``open exchange.''

The core insight is that brand autonomy and cross-brand utility are not mutually exclusive given proper coordination mechanisms. Brands can maintain complete control over their reward programs (earning rates, redemption values, customer relationships, data ownership) while simultaneously enabling customers to exchange rewards across brands through trustless protocol-based mechanisms. This differs fundamentally from traditional coalitions in two critical ways. First, brands issue and manage their own brand-specific rewards rather than generic coalition currency, preserving brand identity. Second, exchange occurs through algorithmic mechanisms embedded in smart contracts rather than through centralized operator intermediation, eliminating rent-seeking and misaligned incentives.

The hybrid architecture addresses each of the nine coalition challenges identified in Section~\ref{subsec:coalition_failures}. Brand identity is preserved because rewards remain brand-specific throughout their lifecycle. Partner cannibalization is mitigated through selective partnership mechanisms where brands control which other brands can accept their rewards. Data ownership remains with brands because customers register with individual brands rather than with central operators. Reputation remains isolated because brands maintain separate identities without shared coalition branding. Complexity is reduced through protocol abstraction where sophisticated exchange mechanisms operate invisibly to users. Economic incentives align because no intermediary extracts rent between brands and customers. Personalization is preserved because brands retain full customer data access. Value and exclusivity are protected through brand control over partnership networks. Economic sustainability improves dramatically because minimal infrastructure costs enable break-even at modest scale.

\begin{figure}[t]
	\centering
	\includegraphics[width=0.5\textwidth]{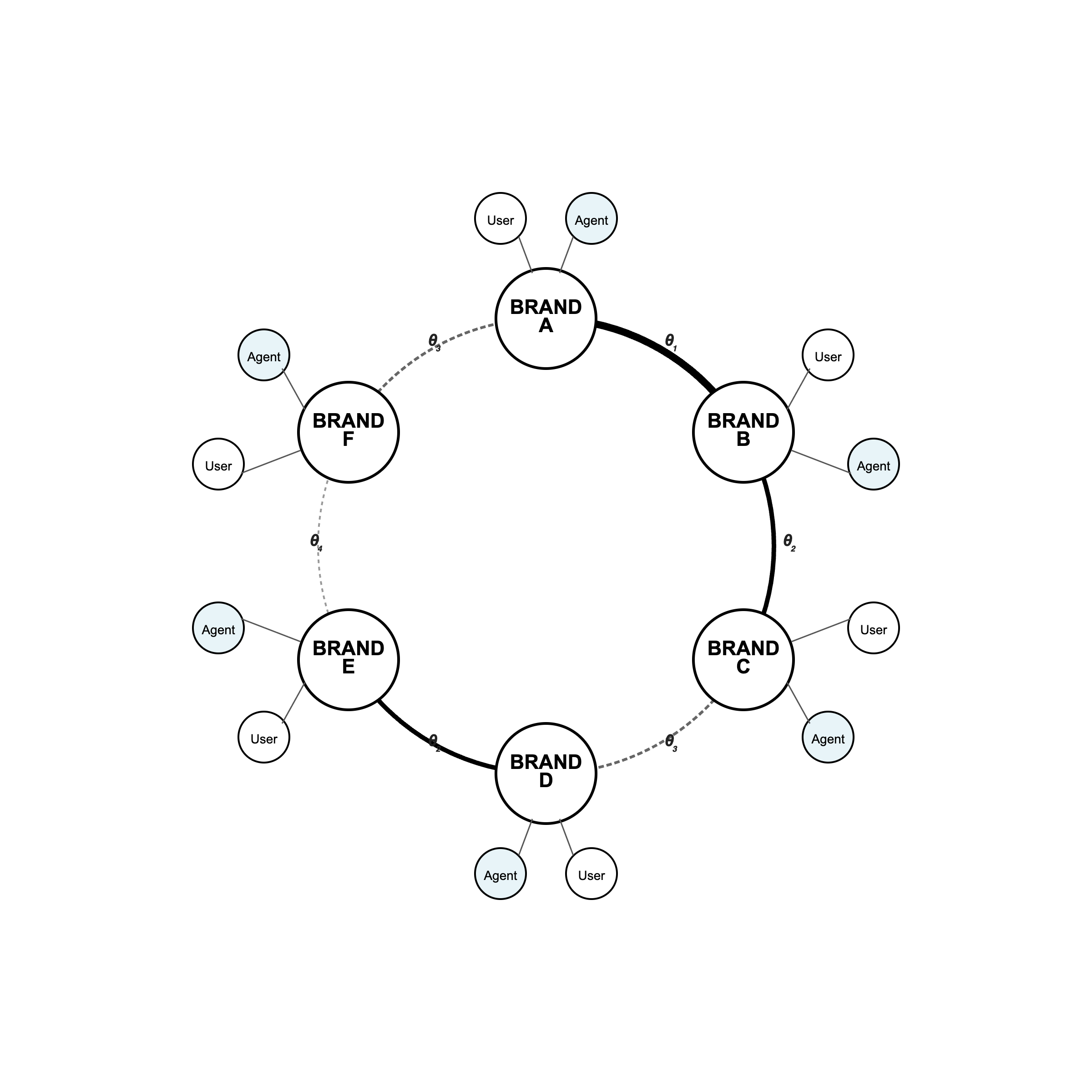}
	\caption{Hybrid Loyalty System Model. Brands maintain sovereign control while participating in a decentralized network with protocol-mediated bilateral flows. The circumference segments represent varying openness parameters: $\theta_1$ (thickest) indicates highest cross-brand exchange openness, decreasing through $\theta_2$, $\theta_3$, to $\theta_4$ (thinnest dashed) representing most restrictive bilateral flows. Each brand serves both human users and AI agents. Users can transact across brands based on openness parameters, though cross-brand flows are not depicted to avoid visual complexity.}
	\label{fig:hybrid_system}
\end{figure}

\subsection{Mathematical Formulation}
\label{subsec:mathematical_formulation}

We formalize the hybrid model by extending closed-system economics with a carefully defined degree of openness. This formalization serves two purposes. It provides precise specification of what makes a system ``hybrid'' rather than purely closed or purely open. And it enables rigorous analysis of how openness affects system properties and outcomes.

\subsubsection{Closed System Model}

A closed loyalty system can be modeled as a function of brand control factors. Let $\Phi_{\text{closed}}$ represent the value that a closed loyalty system delivers to customers and brands. We express this as:
\begin{equation}
	\Phi_{\text{closed}} = \alpha \cdot T + b + \varepsilon
	\label{eq:closed_system}
\end{equation}
where $T$ represents the vector of brand-controlled factors (earning rates, redemption values, tier structures, personalization parameters), $\alpha$ represents the weight vector indicating how each factor influences system value, $b$ captures systematic bias reflecting overall brand premium or discount, and $\varepsilon$ represents random variation from unmodeled factors with $\varepsilon \sim N(0, \sigma^2)$.

The linear form is appropriate for closed systems because brands directly control parameters and factors represent independent decisions that combine additively. A brand improving its earning rate and adding personalization increases value through the sum of these improvements. This model captures how brands like Starbucks or Amazon optimize their closed programs by adjusting controllable parameters.

\subsubsection{Open System Model}

An open coalition system can be similarly modeled as a function of universal coalition parameters. Let $\Phi_{\text{open}}$ represent coalition value:
\begin{equation}
	\Phi_{\text{open}} = \beta \cdot U + c + \varepsilon
	\label{eq:open_system}
\end{equation}
where $U$ represents the vector of coalition-wide factors (universal earning rate, breadth of partner network, redemption options, complexity), $\beta$ represents coalition parameter weights, $c$ captures network effects from coalition membership, and $\varepsilon$ represents random variation.

Again, the linear form is appropriate because coalition governance sets uniform parameters that apply across all partners. Factors represent independent coalition policy decisions whose effects combine additively. A coalition expanding its partner network and improving its redemption catalog increases value through the sum of these enhancements.

\subsubsection{Hybrid Extension}

A hybrid system extends the closed baseline with a degree of openness. Let $\Phi_{\text{hybrid}}$ represent hybrid system value:
\begin{equation}
	\Phi_{\text{hybrid}} = \Phi_{\text{closed}} + \theta
	\label{eq:hybrid_system}
\end{equation}
where $\theta = [P, D]$ represents the degree of openness comprising pricing mechanism $P$ for cross-brand reward exchange and data sharing protocol $D$ for inter-brand data flows.

\subsection{Research Scope: Focus on Pricing Mechanisms}
\label{subsec:research_scope}

The degree of openness $\theta$ comprises pricing mechanism $P$ (determining exchange rates when customers redeem across brands) and data sharing protocol $D$ (governing inter-brand information flows for personalization and attribution) \citep{kumar2016creating}. Both are essential for fully functional hybrid systems, but present distinct research challenges. This paper focuses exclusively on developing $P$, deferring $D$ to future work. Pricing represents the more fundamental challenge because exchange cannot occur without fair valuation, involves complex economic modeling incorporating market dynamics and strategic behavior \citep{tirole1988theory}, and can be evaluated through economic simulation.

The pricing mechanism must address the critical question traditional coalitions solved through centralized operators. How should rewards from different brands be valued relative to each other? In trustless hybrid systems, exchange rates must emerge algorithmically from transparent, verifiable rules that no single actor can manipulate.

Our approach develops a dynamic pricing mechanism accounting for empirically-validated market factors while ensuring fairness, stability, and incentive compatibility. The mechanism operates through dual valuation modes. Same-brand redemption uses the brand's internal valuation where customers redeem directly at brand-controlled rates. Cross-brand redemption uses algorithmic formulas determining exchange rates based on market conditions, brand relationships, transaction characteristics, and system state.

This dual-valuation structure preserves brand autonomy while enabling interoperability. The mechanism must satisfy several properties derived from mechanism design theory \citep{myerson1981optimal, vickrey1961counterspeculation}. Individual rationality requires brands achieve positive expected returns. Fairness demands similar transactions receive similar pricing. Stability requires exchange rates remain within predictable bounds. Budget balance ensures system-wide token conservation \citep{arrow1954existence}. Incentive compatibility prevents gaming opportunities.

The following section derives this pricing mechanism systematically, identifying empirically-validated factors influencing exchange rates, specifying their mathematical combination, and introducing a compensation layer adjusting for externalities not captured by customer-facing prices.

\section{Pricing Mechanism: Mathematical Derivation}
\label{sec:pricing_mechanism}

\subsection{The Fundamental Pricing Challenge}
\label{subsec:pricing_challenge}

When customers redeem rewards across brands, the system must determine fair exchange rates that preserve value for all participants. If a customer holds rewards from Brand A and wishes to redeem them at Brand B, how many Brand A rewards equal the required Brand B rewards? This pricing challenge extends beyond simple proportionality to incorporate market dynamics, strategic relationships, and systemic constraints. We develop a three-layer economic model addressing distinct but interrelated aspects of cross-brand value exchange.

\subsection{Layer One: The Customer Pricing Mechanism}
\label{subsec:layer_one}

The customer pricing mechanism determines what customers pay in source brand rewards to obtain redemptions at destination brands. This customer-facing price must accomplish several objectives beyond simple value transfer, including demand management, behavioral incentives, ecosystem health protection, and perceived fairness.

\subsubsection{Base Proportionality}

We begin with value parity as the foundation. Let $Y$ denote the quantity of destination brand (Brand B) rewards required for the desired redemption. Let $P_A$ and $P_B$ represent the dollar-equivalent redemption values each brand assigns to its own rewards within its store. The base exchange rate establishing value equivalence is:
\begin{equation}
	P_{\text{base}} = Y \times \frac{P_B}{P_A}
	\label{eq:base_proportionality}
\end{equation}

If a customer requires $Y = 20$ Brand B rewards where Brand B values each reward at $P_B = \$0.50$, and Brand A values its rewards at $P_A = \$0.10$, then base proportionality yields $P_{\text{base}} = 20 \times (0.50/0.10) = 100$ Brand A rewards. This baseline ensures customers receive equivalent dollar value whether redeeming at their origin brand or exchanging to redeem elsewhere.

However, market realities make certain exchanges more or less valuable than nominal parity suggests. Loyalty program operators adjust redemption values based on strategic considerations including demand management, competitive positioning, and relationship building \citep{dorotic2014reward}. Our pricing mechanism incorporates similar intelligence through systematic adjustment factors.

\subsubsection{Dynamic Adjustment Factors}

Academic research on coalition loyalty programs and reward economics has identified factors that empirically affect redemption values and program outcomes. We draw from this literature to specify adjustments modifying base pricing to reflect market realities:

\textbf{Cross-Brand Spillover Effects ($\sigma_{AB}$).} Complementary brands create mutual customer value through natural shopping patterns. Research analyzing coalition programs found that popularity affinity represents the main determinant of cross-reward effects, with customers more likely to engage across brands they perceive as naturally paired \citep{schumann2014targeted}. Coffee shops and bakeries exhibit high complementarity, as do airlines and hotels for travelers. When Brand A customers redeem at complementary Brand B, positive spillover enhances overall ecosystem value creation. The spillover factor $\sigma_{AB} \in [0, 1]$ captures this complementarity, where $\sigma_{AB} = 1$ indicates perfect complementarity warranting pricing discounts to encourage beneficial exchanges, and $\sigma_{AB} = 0$ indicates no relationship.

\textbf{Competition and Cannibalization ($\kappa_{AB}$).} When brands compete in overlapping markets, reward exchanges enable customers to leverage partner market power in ways that affect competitive dynamics \citep{meyerwaarden2009grocery}. A customer redeeming coffee shop rewards at a competing coffee chain represents fundamentally different dynamics than redeeming at a complementary bakery. The source brand loses a customer to a rival, incurring costs beyond the nominal transaction value. The cannibalization factor $\kappa_{AB} \in [0, 1]$ reflects competitive intensity, where $\kappa_{AB} = 1$ indicates direct head-to-head competition warranting pricing premiums, and $\kappa_{AB} = 0$ indicates no competitive relationship.

\textbf{Demand Imbalance and Market Saturation ($\delta_B$).} Popular brands experience different cross-reward economics than less popular brands, with popularity driving financial outcomes in coalition settings \citep{liu2009}. High-demand brands accepting rewards during peak periods face opportunity costs from displaced cash-paying customers. Conversely, brands with excess capacity benefit from filling inventory through reward redemptions that stimulate demand during slow periods. The demand imbalance factor $\delta_B \in [-1, +1]$ represents the supply-demand ratio at Brand B, where positive values indicate high demand and scarcity (warranting premiums), and negative values indicate excess capacity (warranting discounts).

\textbf{Temporal and Seasonal Variation ($\xi_B$).} Reward redemption timing affects both customer behavior and brand economics, with empirical evidence showing effects on pre-redemption and post-redemption purchase patterns \citep{dorotic2014reward}. Hotels during summer vacation periods, restaurants during holiday seasons, and retailers during peak shopping windows face higher opportunity costs from reward acceptance than during off-peak periods. The temporal factor $\xi_B \in [-1, +1]$ captures these dynamics, where positive values indicate peak periods with high opportunity costs (warranting premiums), and negative values indicate off-peak periods where brands prefer demand stimulation (warranting discounts).

\textbf{Service Quality and Reputation ($\rho_B$).} Service failures at partner brands negatively affect customer perception through spillover effects, even when the failure is attributable to a specific brand \citep{schumann2014targeted}. While the hybrid architecture maintains more reputation isolation than traditional coalitions, brand quality still signals value in exchange decisions. High-reputation brands may command premiums reflecting their quality positioning, while quality concerns may warrant discounts. The quality factor $\rho_B \in [0, 1]$ represents reputation strength, where $\rho_B = 1$ indicates excellent quality and $\rho_B = 0$ indicates poor quality requiring substantial discounts.

\textbf{Customer Tier Effects ($\Omega$).} Membership tier level significantly affects both attitudinal and behavioral loyalty, with elite members showing substantially higher loyalty scores than base members \citep{kumar2016creating}. Active redeemers demonstrate different spending patterns than enrolled but inactive members. The tier factor $\Omega \in [-\omega_{\max}, +\omega_{\max}]$ aggregates membership tier, redemption history, account tenure, and engagement level to reflect customer lifetime value differentials that warrant pricing adjustments. High-value customers might receive favorable pricing to encourage continued engagement.

\textbf{Flow Balance and Outflow Risk ($\phi_A$).} In coalition programs, brands fund points when customers earn them. When a brand’s customers consistently redeem at partner brands without
attracting corresponding return traffic, the issuing brand experiences a form of cannibalization, having invested in rewards that generate transactions elsewhere \citep{dorotic1999synergistic}. The flow balance factor $\phi_A$ addresses this dynamic by providing pricing signals that discourage excessive outflow when imbalances persist, promoting network reciprocity.

\textbf{Transaction Size and Market Impact ($\mu$).} In liquidity-based exchange systems, large transactions relative to available pool capacity cause disproportionate price impact, affecting subsequent
exchange availability and pricing stability for other participants \citep{angeris2021uniswap}. The transaction size factor $\mu$ addresses this dynamic by enabling pricing adjustments based on relative transaction magnitude. Larger transactions face premiums reflecting their disproportionate impact on system liquidity.


\subsubsection{Multiplicative Pricing Structure}

These factors combine multiplicatively rather than additively because pricing adjustments represent percentage changes that compound. A 20\% discount for spillover effects and a 15\% premium for competition combine as $(1.0 \times 0.80 \times 1.15 = 0.92)$, reflecting how market forces interact multiplicatively rather than independently. Multiplicative structure also ensures scale invariance, where pricing remains proportionally fair regardless of absolute reward values.

The complete customer pricing model expresses the number of Brand A rewards a customer must pay:
\begin{equation}
	P_{\text{customer}} = P_{\text{base}} \times \prod_{i=1}^{n} [1 + \beta_i \times f_i(\Psi_i)] + \varepsilon
	\label{eq:customer_pricing_general}
\end{equation}
where $f_i$ represents the adjustment factor function for factor $i$, $\Psi_i$ represents the factor value, $\beta_i$ represents the parameter weight controlling how strongly factor $i$ influences pricing, and $\varepsilon \sim N(0, \sigma^2_\varepsilon)$ captures unmodeled variation from effects not explicitly tracked. The parameters $\beta$ require empirical calibration as their optimal values depend on specific market contexts.

Expanding Equation~\ref{eq:customer_pricing_general} with our eight identified factors yields:
\begin{align}
	P_{\text{customer}} &= P_{\text{base}} \times \left[1 + \beta_{\text{trans}} \times \max\left(0, \frac{\mu - \eta}{1-\eta}\right)\right] \times \left[1 + \beta_{\text{flow}} \times \phi_A\right] \nonumber \\
	&\quad \times \left[\left(1 - \beta_{\text{spillover}} \times \sigma_{AB}\right) \times \left(1 + \beta_{\text{cannibal}} \times \kappa_{AB}\right)\right] \nonumber \\
	&\quad \times \left[\left(1 + \beta_{\text{demand}} \times \delta_B\right) \times \left(1 + \beta_{\text{season}} \times \xi_B\right)\right] \nonumber \\
	&\quad \times \left[1 + \beta_{\text{quality}} \times (1 - \rho_B)\right] \times [1 + \Omega] + \varepsilon
	\label{eq:customer_pricing_expanded}
\end{align}
where $\eta$ represents a threshold parameter below which transaction size does not trigger adjustments.

\subsubsection{Alternative Representation: Premium and Discount Factors}

The multiplicative form in Equation~\ref{eq:customer_pricing_expanded} can be conceptually reorganized to highlight which factors increase versus decrease customer costs. We can express pricing adjustments as the sum of premium factors minus discount factors:
\begin{equation}
	P_{\text{customer}} = P_{\text{base}} \times \left[1 + \sum_{j \in \text{Premiums}} \beta_j \times f_j(\Psi_j) - \sum_{k \in \text{Discounts}} \beta_k \times f_k(\Psi_k)\right] + \varepsilon
	\label{eq:premium_discount}
\end{equation}

where:

\textbf{Premium Factors} (increase customer cost):
\begin{itemize}
	\item Competition/Cannibalization: $\beta_{\text{cannibal}} \times \kappa_{AB}$
	\item Demand Imbalance (scarcity): $\beta_{\text{demand}} \times \delta_B$ when $\delta_B > 0$
	\item Seasonal Premium (peak periods): $\beta_{\text{season}} \times \xi_B$ when $\xi_B > 0$
	\item Transaction Size (large): $\beta_{\text{trans}} \times \max(0, (\mu-\eta)/(1-\eta))$
	\item Flow Imbalance (excessive outflows): $\beta_{\text{flow}} \times \phi_A$ when $\phi_A > 0$
	\item Quality Discount (poor reputation): $\beta_{\text{quality}} \times (1 - \rho_B)$ when $\rho_B <$ threshold
\end{itemize}

\textbf{Discount Factors} (decrease customer cost):
\begin{itemize}
	\item Spillover/Complementarity: $\beta_{\text{spillover}} \times \sigma_{AB}$
	\item Demand Excess (overcapacity): $\beta_{\text{demand}} \times |\delta_B|$ when $\delta_B < 0$
	\item Seasonal Discount (off-peak): $\beta_{\text{season}} \times |\xi_B|$ when $\xi_B < 0$
	\item Customer Tier Benefits: $\Omega$ when $\Omega > 0$
	\item Flow Balance Incentive: $\beta_{\text{flow}} \times |\phi_A|$ when $\phi_A < 0$
\end{itemize}

This representation clarifies the economic intuition underlying the pricing mechanism. Premiums discourage exchanges that harm ecosystem health or impose hidden costs. Discounts encourage exchanges that benefit the ecosystem or occur under favorable conditions. The net adjustment reflects the balance of these competing forces for any specific exchange.


\subsection{Layer Two: The Inter-Brand Settlement Challenge}
\label{subsec:layer_two}

The pricing mechanism in Section~\ref{subsec:layer_one} determines what customers pay. However, this addresses only the customer-facing aspect of cross-brand exchange. A deeper challenge emerges when we consider the mechanics of actual value transfer between brands. How does value flow from Brand A to Brand B to compensate for the service Brand B provides?

\subsubsection{The Sovereignty Constraint}

Our hybrid architecture establishes a foundational design principle that distinguishes it from traditional coalitions. Each brand issues, manages, and accepts only its own brand-specific rewards. Starbucks accepts Starbucks Stars. The bakery accepts bakery points. Brands do not directly accept rewards from other brands. This sovereignty is not an arbitrary restriction but an essential feature preserving brand identity and eliminating the centralized dependencies that doom traditional coalitions.

Consider what occurs in a cross-brand redemption. A customer presents Brand A rewards. Brand B must provide service or product at a discount. But Brand B only recognizes Brand B rewards as valid for redemption within its store. The customer's Brand A rewards have no direct meaning or value to Brand B. This creates a puzzle. The customer pays Brand A rewards. Brand B requires compensation for providing service. Brand B will not accept Brand A rewards as payment. How does value flow from Brand A to Brand B given this constraint?

\subsubsection{The Naive Broker Architecture}

A natural solution emerges from considering international currency exchange. When travelers need euros but hold dollars, currency exchange services facilitate conversion. These services maintain inventories of multiple currencies and execute exchanges at prevailing rates. We might envision an analogous mechanism for reward exchange.

Introduce a broker entity that maintains inventories of rewards from all participating brands. When a customer seeks to redeem Brand A rewards at Brand B, the broker accepts Brand A rewards from the customer, determines equivalent value in Brand B rewards, and provides those Brand B rewards to Brand B. Brand B receives the reward currency it recognizes as valid. The customer achieves cross-brand redemption. Brand A's rewards exit circulation.

For this to function across a network of $N$ brands, the broker must hold meaningful inventories of rewards from Brand A, Brand B, Brand C, through Brand N. These inventories determine system capacity. The broker cannot facilitate an exchange requiring 100 Brand B rewards if it possesses only 50. The broker must acquire these reward inventories from brands through some deposit or liquidity provision mechanism.

\subsubsection{The Verification Problem}

This broker architecture encounters a fundamental difficulty that proves insurmountable under our design constraints. To determine exchange rates between different brand rewards, the broker must know the value of each brand's rewards in some comparable unit.

Brands assign redemption values to their rewards within their own stores. Brand A might specify that 100 Brand A rewards redeem for a \$10 product, implying $P_A^{\text{store}} = \$0.10$ per reward. Brand B might offer redemptions at $P_B^{\text{store}} = \$0.50$ per reward. The broker might naturally use these store-assigned values to determine exchange rates, concluding that 5 Brand A rewards equal 1 Brand B reward based on their respective \$0.10 and \$0.50 values.

However, consider a critical challenge. How does the broker verify these claimed values in a protocol designed to operate among untrusted brands? A malicious or mistaken brand could claim arbitrary reward values. Brand C might claim $P_C^{\text{store}} = \$1{,}000$ per reward and list some product at this inflated value, creating apparent internal consistency. But the claimed value bears no relationship to actual economic value. The brand manufactures arbitrary numbers.

In traditional coalitions, central operators maintain contractual relationships enabling audits, facility visits, product offering verification, and enforcement through reputation and contract terms. But our hybrid model explicitly eliminates centralized operators to avoid rent-seeking and misaligned incentives. We are building a protocol for coordination among potentially adversarial or at minimum mutually distrustful brands. No trusted third party exists who can validate each brand's claimed reward value. The broker cannot simply trust what brands report.

This verification impossibility creates a severe constraint. If the broker cannot trust claimed values and cannot verify them independently at scale, the broker must either reject unvalidated brands or risk accepting fraudulent valuations. The former leads directly back to centralized control where some authority decides who participates. The latter makes the system exploitable and economically incoherent.

\subsubsection{The Universal Asset Solution}

We resolve this through an elegant mechanism inspired by commodity-backed monetary systems. Rather than relying on unverifiable reward value claims, we introduce a universal medium of exchange that all brands objectively recognize as valuable. We denote this universal asset as $M$.

The essential property $M$ must possess is universal acceptance. Every brand in the system must both understand the value of $M$ and willingly accept $M$ as legitimate payment. Consider the role of national currency in domestic commerce. Every business understands what a dollar represents. Every business accepts dollars as payment. This universal acceptance eliminates verification requirements. Merchants need not trust customers' personal value assessments. The currency itself carries objective, collectively recognized worth.

For our reward exchange protocol, $M$ serves an analogous function. We require that $M$ be an asset with transparent, verifiable value that all participating brands already accept in normal business operations. This might be fiat currency itself, or in modern digital contexts, a stablecoin asset with transparent market pricing. The specific nature of $M$ matters less than its universal acceptance property. Every brand agrees that $M$ has definite value and willingly accepts $M$ as payment.

\subsubsection{The Backing Mechanism}

The universal asset $M$ resolves our verification problem through economic commitment rather than trust. Instead of brands depositing rewards and making unverifiable value claims, we require brands to back their reward deposits with equivalent value in the universal asset $M$.

Suppose Brand A wishes to participate in cross-brand exchange, making its rewards redeemable at other brands. Brand A must deposit some quantity $X_A$ of Brand A rewards with the broker (or more precisely, with the protocol mechanism that replaces the broker). However, Brand A cannot simply deposit these rewards claiming arbitrary value. Brand A must simultaneously deposit $M$ assets in an amount reflecting the value Brand A assigns to these $X_A$ rewards.

Let $P_A$ represent the dollar-equivalent value Brand A assigns per reward as established in Equation~\ref{eq:base_proportionality}. Let $P_M$ represent the value per unit of asset $M$ (typically $P_M = \$1$ if $M$ is dollar-denominated). If Brand A deposits $X_A$ rewards valued at $P_A$ each, Brand A must deposit:
\begin{equation}
	M_A^{\text{deposit}} = \frac{X_A \times P_A}{P_M}
	\label{eq:m_deposit}
\end{equation}
units of universal asset $M$ alongside these rewards. This $M$ deposit testifies to the claimed reward value through verifiable economic commitment. Brand A cannot claim its rewards are worth \$1,000 each without depositing $\$1{,}000 \times X_A$ worth of $M$ assets. The $M$ backing transforms unverifiable claims into observable commitments.

The settlement mechanism now operates on $M$ assets rather than attempting to verify reward values directly. When facilitating an exchange, the system does not ask ``what are these rewards worth?'' but rather observes ``what $M$ assets back these rewards?'' This observable quantity $M_A^{\text{deposit}} / X_A$ provides objective per-reward valuation $P_A$ without requiring trust or verification of claims.


\subsubsection{The Settlement Mechanism}

Consider how settlement operates under $M$-backed architecture. A customer holds Brand A rewards and seeks redemption at Brand B requiring $Y$ Brand B rewards. We work systematically through the settlement process:

\textbf{Step 1: Identify Destination Requirements.} Brand B's redemption structure specifies that the customer needs $Y = 20$ Brand B rewards to obtain the desired product or discount. This is determined by Brand B's pricing and is independent of cross-brand mechanics.

\textbf{Step 2: Calculate Required $M$ Value.} Since Brand B deposited its rewards backed by $M$ assets at value $P_B$ per reward, $Y$ Brand B rewards correspond to $M$ value:
\begin{equation}
	M_{\text{required}} = Y \times P_B
	\label{eq:m_required}
\end{equation}
If $P_B = \$0.50$ per Brand B reward and $Y = 20$ rewards are needed, then $M_{\text{required}} = 20 \times 0.50 = \$10$. This is the value that must flow from Brand A's account to Brand B's account to compensate Brand B fairly for providing the redemption.

\textbf{Step 3: Execute $M$ Asset Transfer.} The protocol transfers $M_{\text{required}} = \$10$ worth of $M$ assets from Brand A's deposited pool to Brand B's deposited pool. This settlement occurs at base economic value regardless of customer pricing adjustments. Brand B provides \$10 worth of value through the discount. Brand B must receive \$10 worth of value as compensation. This is fair value exchange.

\textbf{Step 4: Release Destination Rewards.} Having received $M$ asset compensation of $M_{\text{required}}$, the protocol releases $Y = 20$ Brand B rewards from Brand B's deposited inventory. These rewards flow to Brand B's redemption system where the customer's transaction is processed according to Brand B's standard procedures.

\textbf{Step 5: Accept Customer Payment.} The customer provides $P_{\text{customer}}$ Brand A rewards as determined by Equation~\ref{eq:customer_pricing_expanded}. These rewards are removed from circulation and credited to Brand A's deposited pool. Note that $P_{\text{customer}}$ may differ from the base proportionality value due to market factor adjustments, but the settlement in Step 3 transfers $M$ value at base proportionality regardless.

\subsubsection{Settlement at Base Value}

The critical insight is that settlement always occurs at base value using $M$ assets, independent of customer pricing adjustments. Regardless of whether the customer pays 80, 100, or 120 Brand A rewards due to premiums or discounts, the settlement layer transfers exactly $M_{\text{required}} = Y \times P_B$ in $M$ assets from Brand A to Brand B.

This separation serves essential functions. Customer pricing manages demand, incentivizes beneficial exchanges, and discourages harmful ones through market-sensitive adjustments. Settlement ensures brands always receive fair value for services provided, maintaining economic integrity independent of behavioral incentive mechanisms. These two functions operate independently, enabling optimization along different dimensions simultaneously.

The settlement transfer for any transaction is:
\begin{equation}
	S_M = Y \times P_B
	\label{eq:settlement_transfer}
\end{equation}
flowing from source brand to destination brand in universal asset $M$.

\subsubsection{Emergent Properties: Decentralized Control of Openness}

The $M$-backed settlement architecture yields remarkable properties addressing several challenges simultaneously.

\textbf{Brand B Controls Reception Capacity.} Brand B deposited some quantity $X_B$ of Brand B rewards with the protocol, backed by $M_B = X_B \times P_B$ in $M$ assets. Every cross-brand redemption directed toward Brand B consumes $Y$ rewards from this inventory per Equation~\ref{eq:settlement_transfer}. When Brand B's deposited reward inventory is exhausted, the protocol cannot facilitate additional exchanges toward Brand B, even if customers possess sufficient source brand rewards and wish to make such exchanges. The system automatically rejects requests because it lacks the necessary Brand B rewards to complete settlement.

This creates an elegant result. Brand B controls the volume of cross-brand redemptions it accepts through the quantity of rewards it deposits with the protocol. If Brand B wishes to accept only limited cross-brand traffic, it deposits a small inventory. If Brand B wishes to encourage substantial cross-brand redemptions, it deposits a large inventory. Brand B modulates its degree of openness through a simple operational parameter under its complete control, without requiring permission from central authorities or coordination with other brands. This preserves brand sovereignty while enabling interoperability.

\textbf{Brand A Controls Outbound Capacity.} Symmetrically, Brand A's customers can only redeem at other brands to the extent Brand A has deposited $M$ assets backing its rewards. Each cross-brand redemption transfers $M$ assets from Brand A's pool to destination brands per Equation~\ref{eq:settlement_transfer}. When Brand A's $M$ asset balance is depleted to levels insufficient to back remaining Brand A rewards, its customers can no longer redeem elsewhere, even if destination brands have available inventory. Brand A controls how much cross-brand utility it provides through its $M$ asset deposits, again without requiring external permission or coordination.

\textbf{Objective Valuation Without Trust.} The $M$ backing eliminates verification requirements entirely. The protocol need not trust Brand A's claims about $P_A$. The protocol need not investigate Brand A's business to verify product pricing. The protocol simply observes that Brand A deposited $X_A$ rewards backed by $M_A^{\text{deposit}}$ assets, implying $P_A = M_A^{\text{deposit}} / X_A$ as the verifiable value per reward. This value reflects Brand A's own economic commitment rather than an unverifiable claim. If Brand A inflates claimed values, Brand A must deposit correspondingly large $M$ assets, making fraudulent inflation economically expensive.

\textbf{Automated Trustless Operation.} Because all relevant information exists in observable quantities ($M$ deposits, reward inventories, exchange requests), the settlement mechanism can operate algorithmically without human judgment or discretion. This enables implementation through automated market makers and smart contracts, eliminating centralized broker entities while maintaining functional coordination. The system becomes trustless and perpetual, operating according to programmed rules without dependence on any party's continued operation or good faith.


\subsection{Layer Three: The Compensation Mechanism}
\label{subsec:layer_three}

The pricing mechanism (Layer 1) determines what customers pay. The settlement mechanism (Layer 2) ensures brands receive fair base value for services provided. However, certain market dynamics create asymmetric costs and benefits not fully captured by either customer pricing or base value settlement. These externalities require additional inter-brand adjustments to maintain economic equilibrium and ensure all parties achieve fair outcomes.

\subsubsection{The Limitations of Settlement Alone}

Consider a cross-brand redemption where Brand A and Brand B are direct competitors. The customer pays $P_{\text{customer}}$ Brand A rewards per the pricing mechanism. Brand B receives $M_{\text{required}} = Y \times P_B$ in settlement per Equation~\ref{eq:settlement_transfer}. This settlement compensates Brand B for the nominal value of service provided. However, Brand A suffers additional costs beyond this base value transfer.

Brand A loses the customer to a competitor. This customer might have made future purchases at Brand A but may now shift loyalty toward Brand B. Brand A incurred customer acquisition costs to establish this customer relationship, which are now lost. Brand A sacrifices future revenue streams from this customer's potential repeat business. These competitive damages exceed the nominal transaction value settled through $M$ asset transfer.

Symmetrically, Brand B gains a new customer without incurring typical customer acquisition costs. Brand B benefits from Brand A's marketing efforts and relationship building. This competitive advantage creates value for Brand B beyond the base value settlement received. Fair economic allocation suggests Brand B should compensate Brand A for this customer relationship transfer and competitive damage.

\subsubsection{Temporal Opportunity Costs}

Consider a different scenario where Brand B operates a restaurant accepting redemptions during peak dining hours. Brand B receives $M_{\text{required}}$ settlement per Equation~\ref{eq:settlement_transfer}. However, accepting reward redemptions means displacing potential cash-paying customers. If Brand B is capacity-constrained and operating at maximum seating, every table occupied by a reward-redeeming customer represents a table unavailable for a cash customer who might have paid full price. The opportunity cost of this displaced revenue exceeds the base value settlement Brand B receives.

Conversely, during off-peak hours when the restaurant operates below capacity, accepting reward redemptions provides value to Brand B beyond the settlement received. These redemptions fill otherwise empty capacity, stimulate demand during slow periods, and may generate ancillary revenue from add-on purchases. Brand B benefits from demand stimulation, suggesting Brand B should provide compensation to encourage off-peak redemptions.

\subsubsection{Spillover Value Capture}

Consider a third scenario where Brand A and Brand B are highly complementary businesses such as a coffee shop and bakery located near each other. When Brand A's customer redeems at Brand B, positive spillover may occur. The customer discovers Brand B through the reward redemption. This customer may become a regular Brand B customer even absent future cross-brand redemptions. The customer may simultaneously patronize both businesses in future visits, buying coffee at Brand A and pastries at Brand B during the same shopping trip.

Brand B gains customer acquisition value from this spillover. Brand A provided the referral through its rewards program and customer relationship. Fair economic allocation suggests Brand B should compensate Brand A for this referral value based on the customer's expected lifetime value to Brand B and the strength of spillover effects.

\subsubsection{The Compensation Layer Design}

These externalities operate at the brand relationship level rather than the customer transaction level. Customer pricing influences customer behavior and demand patterns. Settlement ensures fair transactional value transfer. Compensation addresses strategic implications that manifest over time and across repeated interactions. We introduce direct brand-to-brand compensation flows that operate invisibly to customers, adjusting inter-brand settlements to account for these externalities.

The compensation amount flowing between brands in $M$ assets is:
\begin{equation}
	C_{AB} = S_M \times \sum_{i} \gamma_i \times g_i(\text{factor}_i)
	\label{eq:compensation_general}
\end{equation}
where $S_M$ represents the base settlement value from Equation~\ref{eq:settlement_transfer}, $\gamma_i$ represents compensation parameters distinct from customer pricing parameters $\beta_i$, and $g_i$ represents factor functions that may differ from customer pricing factor functions $f_i$. The sum aggregates across compensation-relevant factors.

\subsubsection{Competition Compensation}

When customer redemptions flow from Brand A to competing Brand B, Brand B gains customers at Brand A's expense. The compensation parameter $\gamma_{\text{cannibal}}$ determines how much Brand B pays Brand A to compensate for competitive damage. This should reflect customer acquisition costs, expected customer lifetime value, and competitive relationship intensity $\kappa_{AB}$.
\begin{equation}
	C_{\text{competition}} = S_M \times \gamma_{\text{cannibal}} \times \kappa_{AB}
	\label{eq:comp_competition}
\end{equation}
where $\kappa_{AB}$ is the competition factor from Section~\ref{subsec:layer_one}. Higher competition intensity warrants larger compensation flows. Direct competitors ($\kappa_{AB} \approx 1$) require substantial compensation. Brands in different markets ($\kappa_{AB} \approx 0$) require negligible compensation.

\subsubsection{Seasonal Opportunity Cost Compensation}

When Brand B accepts redemptions during peak periods with high opportunity costs, Brand B deserves compensation beyond base settlement for displaced revenue. Conversely, when Brand B benefits from off-peak redemptions stimulating demand during excess capacity, Brand B should provide negative compensation (effectively paying Brand A for the demand).
\begin{equation}
	C_{\text{seasonal}} = S_M \times \gamma_{\text{season}} \times \xi_B
	\label{eq:comp_seasonal}
\end{equation}
where $\xi_B$ is the temporal factor from Section~\ref{subsec:layer_one}. Positive $\xi_B$ during peak periods results in positive compensation flowing to Brand B. Negative $\xi_B$ during off-peak periods results in negative compensation (payment from Brand B to Brand A).

\subsubsection{Spillover Value Compensation}

When Brand B gains valuable customers through positive spillover from Brand A, Brand B should compensate Brand A for referral value. This compensation should reflect customer lifetime value estimates and spillover strength $\sigma_{AB}$.
\begin{equation}
	C_{\text{spillover}} = S_M \times \gamma_{\text{synergy}} \times \sigma_{AB} \times \text{LTV}_B
	\label{eq:comp_spillover}
\end{equation}
where $\sigma_{AB}$ is the spillover factor from Section~\ref{subsec:layer_one} and $\text{LTV}_B$ represents estimated customer lifetime value at Brand B. Strong complementarity ($\sigma_{AB} \approx 1$) with high customer value (large $\text{LTV}_B$) warrants substantial compensation flowing from Brand B to Brand A.

\subsubsection{Net Compensation Flow}

The total compensation combines these components:
\begin{equation}
	C_{AB} = C_{\text{competition}} + C_{\text{seasonal}} + C_{\text{spillover}}
	\label{eq:compensation_total}
\end{equation}

Positive $C_{AB}$ indicates net compensation flows from Brand B to Brand A. Negative $C_{AB}$ indicates net compensation flows from Brand A to Brand B. Compensation occurs in $M$ assets directly between brand pools, operating independently of customer-facing transactions.

\subsubsection{Parameter Calibration}

The compensation parameters $\gamma_i$ require empirical calibration distinct from customer pricing parameters $\beta_i$. Customer pricing parameters $\beta_i$ optimize for demand management and customer incentives. Compensation parameters $\gamma_i$ optimize for inter-brand equity and sustainable economics. These serve different objectives and warrant independent calibration based on observed brand behavior, participation patterns, and system stability metrics.


\subsection{The Complete Three-Layer Transaction Flow}
\label{subsec:complete_flow}

A complete cross-brand redemption orchestrates flows across all three layers simultaneously within a single atomic transaction. Understanding how these layers interact provides insight into the system's economic elegance and operational coherence.

\subsubsection{Transaction Example}

Consider a concrete scenario. A customer holds Brand A (coffee shop) rewards and seeks a \$10 discount at Brand B (bakery). Brand B requires $Y = 20$ Brand B rewards for this discount, where each Brand B reward is valued at $P_B = \$0.50$. Brand A values its rewards at $P_A = \$0.10$. The brands are complementary ($\sigma_{AB} = 0.8$), not competitive ($\kappa_{AB} = 0.1$), and the redemption occurs during Brand B's peak season ($\xi_B = 0.5$). The customer is a high-tier member ($\Omega = 0.05$).

\textbf{Layer 1: Customer Pricing Calculation.} Using Equation~\ref{eq:base_proportionality}, base proportionality yields:
\[
P_{\text{base}} = 20 \times \frac{0.50}{0.10} = 100 \text{ Brand A rewards}
\]

Applying dynamic factors from Equation~\ref{eq:customer_pricing_expanded} (simplified for illustration):
\begin{itemize}
	\item Spillover discount: $(1 - 0.3 \times 0.8) = 0.76$ (24\% discount for high complementarity)
	\item Competition premium: $(1 + 0.2 \times 0.1) = 1.02$ (2\% premium for minor competition)
	\item Seasonal premium: $(1 + 0.3 \times 0.5) = 1.15$ (15\% premium for peak season)
	\item Customer tier benefit: $(1 + 0.05) = 1.05$ (5\% discount for high-tier member)
\end{itemize}

Combined adjustment: $0.76 \times 1.02 \times 1.15 \times 1.05 \approx 0.935$
\[
P_{\text{customer}} = 100 \times 0.935 = 93.5 \approx 94 \text{ Brand A rewards}
\]

The customer pays 94 Brand A rewards despite base value requiring 100, reflecting the net effect of discounts (spillover, tier) exceeding premiums (competition, seasonal).

\textbf{Layer 2: Settlement Transfer Calculation.} From Equation~\ref{eq:settlement_transfer}, settlement transfers base value regardless of customer pricing:
\[
S_M = 20 \times 0.50 = \$10 \text{ in } M \text{ assets}
\]

The protocol transfers exactly \$10 $M$ from Brand A's pool to Brand B's pool. This occurs independent of the 94 Brand A rewards the customer paid.

\textbf{Layer 3: Compensation Calculation.} Using Equations~\ref{eq:comp_competition}--\ref{eq:comp_spillover}:

Competition compensation (flows B $\rightarrow$ A):
\[
C_{\text{competition}} = 10 \times 0.15 \times 0.1 = \$0.15
\]

Seasonal compensation (flows B $\leftarrow$ A due to negative sign):
\[
C_{\text{seasonal}} = 10 \times 0.20 \times 0.5 = \$1.00
\]

Spillover compensation (flows B $\rightarrow$ A):
\[
C_{\text{spillover}} = 10 \times 0.10 \times 0.8 = \$0.80
\]

Net compensation from Equation~\ref{eq:compensation_total}:
\[
C_{AB} = 0.15 - 1.00 + 0.80 = -\$0.05
\]

Negative value indicates Brand A pays Brand B an additional \$0.05 in net compensation, primarily because seasonal opportunity costs (\$1.00) exceed competitive and spillover benefits (\$0.95).

\subsubsection{Complete Transaction Settlement}

The atomic transaction executes all three layers:

\textbf{Brand A's Position:}
\begin{itemize}
	\item Receives: 94 Brand A rewards from customer (removed from circulation)
	\item Pays: \$10 $M$ to Brand B (settlement)
	\item Pays: \$0.05 $M$ to Brand B (net compensation)
	\item Net outflow: \$10.05 $M$ for facilitating customer redemption
\end{itemize}

\textbf{Brand B's Position:}
\begin{itemize}
	\item Receives: \$10 $M$ from Brand A (settlement)
	\item Receives: \$0.05 $M$ from Brand A (net compensation)
	\item Provides: 20 Brand B rewards for redemption (service to customer)
	\item Net inflow: \$10.05 $M$ for providing service
\end{itemize}

\textbf{Customer's Position:}
\begin{itemize}
	\item Pays: 94 Brand A rewards
	\item Receives: \$10 discount at Brand B
	\item Net benefit: Cross-brand redemption flexibility
\end{itemize}

The three-layer structure ensures economic coherence. Brand A's total outflow (\$10.05) compensates Brand B for both base value (\$10 settlement) and externalities (\$0.05 net compensation). Customer pricing (94 rewards) manages demand and incentives independently of inter-brand settlements. All flows occur atomically within a single transaction, ensuring consistency.

\subsection{Model Properties and Limitations}
\label{subsec:model_properties}

\subsubsection{Desirable Properties}

The three-layer pricing model exhibits several properties essential for sustainable operation:

\textbf{Individual Rationality.} With proper parameter calibration, brands achieve positive expected returns from participation. Settlement ensures fair value recovery. Compensation adjusts for externalities. Customer pricing generates revenue exceeding costs over time.

\textbf{Fairness.} Similar transactions receive similar treatment through consistent application of factors. Brands in similar competitive positions face similar pricing adjustments. Customers with similar characteristics receive comparable pricing.

\textbf{Objective Valuation.} $M$ asset backing provides verifiable valuation without trust requirements. No party can claim arbitrary values without corresponding economic commitment. Fraud becomes expensive rather than costless.

\textbf{Bounded Pricing.} Factor values and parameters are bounded, ensuring exchange rates remain within predictable ranges. Customers face stable pricing without extreme volatility. Brands can anticipate costs and revenues.

\textbf{Decentralized Control.} Brands independently control their degree of openness through deposit decisions. No central authority approves participation. No operator extracts rent. Protocol mechanisms coordinate automatically.

\textbf{Separation of Concerns.} Customer pricing, settlement, and compensation optimize for distinct objectives independently. Changes to customer incentives do not disrupt inter-brand settlements. Compensation adjustments do not affect customer experience.

\subsubsection{Acknowledged Limitations}

Important limitations warrant recognition:

\textbf{Factor Incompleteness.} The eight identified factors provide first-order approximations. Additional effects including psychological attachments, cultural preferences, network effects beyond direct spillover, and external market shocks influence actual outcomes but are not explicitly modeled. The error term $\varepsilon$ in Equation~\ref{eq:customer_pricing_general} acknowledges unmodeled variation.

\textbf{Measurement Challenges.} Subjective factors such as brand affinity ($\sigma_{AB}$), service quality ($\rho_B$), and spillover effects resist precise quantification. Estimates contain uncertainty. Measurement errors propagate through the pricing model.

\textbf{Dynamic Markets.} Factor values evolve over time. Competition intensifies or relaxes. Demand patterns shift seasonally and cyclically. Quality reputations change. The model requires real-time parameter updates to maintain accuracy.

\textbf{Strategic Behavior.} Brands may attempt to game factor values. A brand might temporarily inflate quality signals before extracting value. Brands might collude to manipulate compensation parameters. Governance mechanisms must detect and prevent gaming.

\textbf{Calibration Requirements.} Parameters $\beta_i$ and $\gamma_i$ require empirical calibration using real transaction data. Optimal values depend on specific market contexts. Initial deployments operate with estimated parameters requiring refinement through observed outcomes.

\textbf{First-Order Approximation.} The model provides tractable approximations rather than complete representations. Simplifications include single-period analysis, static optimal valuations, and linear factor functions. Extensions might incorporate multi-period dynamics, adaptive learning, and nonlinear relationships.

These limitations suggest the model provides a rigorous foundation for implementation while acknowledging that real-world deployment requires continuous validation, parameter estimation updates, and adaptive refinement based on observed system behavior and emerging patterns.


\subsection{System Efficiency and Economic Sustainability}
\label{subsec:system_efficiency}

The $M$-backed settlement architecture might initially appear to impose significant capital requirements on participating brands. Each brand must deposit $M$ assets to back its rewards, creating what seems like a substantial upfront cost. Moreover, as customers redeem Brand A rewards at other brands, $M$ assets flow out of Brand A's pool, suggesting ongoing replenishment requirements that could make participation economically prohibitive for smaller brands. However, deeper analysis reveals an elegant property where $M$ assets circulate within the ecosystem rather than being consumed, transforming apparent costs into measures of system health and efficiency.

\subsubsection{The Circulation Property}

Consider Brand A's $M$ asset flows over time. When Brand A customers redeem at other brands, $M$ assets transfer out of Brand A's pool per Equation~\ref{eq:settlement_transfer}. Let $O_A(t)$ represent Brand A's $M$ asset outflow rate at time $t$, measured in dollars per unit time. These outflows represent value leaving Brand A's pool to compensate other brands for services their customers receive.

However, Brand A simultaneously experiences inflows when customers from other brands redeem at Brand A. Each such redemption transfers $M$ assets from the source brand's pool to Brand A's pool. Let $I_A(t)$ represent Brand A's $M$ asset inflow rate at time $t$. These inflows represent value Brand A receives as compensation for services it provides to others' customers.

The net $M$ asset flow for Brand A is:
\begin{equation}
	N_A(t) = I_A(t) - O_A(t)
	\label{eq:net_flow}
\end{equation}

Positive net flow indicates Brand A is accumulating $M$ assets. Negative net flow indicates Brand A is depleting $M$ assets. However, in a balanced ecosystem where cross-brand redemptions flow roughly symmetrically, we expect $I_A(t) \approx O_A(t)$ over time, yielding $N_A(t) \approx 0$.

\subsubsection{The Equilibrium Insight}

Consider what equilibrium means economically. If Brand A attracts as many redemptions from other brands' customers as its own customers redeem elsewhere, then $M$ assets flow into Brand A at approximately the same rate they flow out. Brand A receives $M$ assets as compensation for services provided (inflows) and pays $M$ assets as compensation for services its customers receive elsewhere (outflows). At equilibrium, these roughly balance.

In this balanced state, Brand A requires no additional $M$ asset deposits beyond its initial commitment. The $M$ assets circulate within the ecosystem, flowing from brand to brand as customers make cross-brand redemptions, but aggregate $M$ quantity remains constant. Brand A's initial deposit serves not as consumed capital but as working capital facilitating transactions. The deposit enables participation but does not represent ongoing costs.

This transforms the economic model from one requiring recurring capital expenditure to one requiring initial capital commitment that remains productive indefinitely. The $M$ assets backing Brand A's rewards continue to facilitate exchanges over time without requiring replenishment, provided the ecosystem maintains approximate balance.

\subsubsection{Depth Requirements and Variance Buffering}

While equilibrium suggests zero net flow over time, short-term flows exhibit variance around this equilibrium. Some periods may see more outflows than inflows. Other periods reverse this pattern. Brand A must maintain sufficient $M$ asset depth to buffer this variance without depleting reserves.

Let $\sigma_A$ represent the standard deviation of Brand A's net flow $N_A(t)$ over some time window. Let $T$ represent the time horizon over which Brand A wishes to maintain operations without requiring additional deposits. Standard buffer stock theory suggests the optimal depth $d_A$ that Brand A should deposit satisfies:
\begin{equation}
	d_A \geq k \cdot \sigma_A \cdot \sqrt{T}
	\label{eq:buffer_depth}
\end{equation}
where $k$ represents a confidence parameter. Choosing $k = 2$ provides approximately 95\% confidence that reserves will not deplete over horizon $T$. Choosing $k = 3$ provides approximately 99.7\% confidence.

This reveals that depth requirements scale with the square root of time horizon and linearly with flow variance. Brands with stable, predictable flows require less depth than brands with volatile flows. Brands planning longer operational horizons without replenishment require more depth. However, depth requirements grow sublinearly with time horizon due to the $\sqrt{T}$ term, reflecting mean reversion properties in balanced systems.

\subsubsection{System-Wide Efficiency Metrics}

The equilibrium concept extends from individual brands to system-wide efficiency. A healthy ecosystem exhibits balanced flows across all participants, minimizing net accumulation or depletion at any individual brand. We can formalize this through a system efficiency metric $\eta$:
\begin{equation}
	\eta = 1 - \frac{\sum_{i=1}^{N} |I_i - O_i|}{\sum_{i=1}^{N} (I_i + O_i)}
	\label{eq:system_efficiency}
\end{equation}
where the sums aggregate across all $N$ brands in the ecosystem over some measurement period. The numerator captures total imbalance (sum of absolute net flows). The denominator captures total activity (sum of all inflows and outflows). The ratio measures what fraction of total flow represents imbalance rather than circulation.

Perfect efficiency $\eta = 1$ occurs when $I_i = O_i$ for all brands, indicating complete balance with zero net flows. Lower efficiency indicates persistent imbalances where some brands accumulate while others deplete. System efficiency below some threshold (say $\eta < 0.7$) might trigger rebalancing mechanisms or signal ecosystem dysfunction requiring intervention.

This metric provides both diagnostic and predictive value. High efficiency indicates sustainable operation requiring minimal $M$ asset replenishment. Low efficiency predicts future capital requirements as depleted brands need additional deposits. Monitoring $\eta$ over time reveals whether the ecosystem is converging toward balance or diverging toward instability.

\subsubsection{Revenue Recognition and Economic Interpretation}

Received $M$ assets represent genuine economic value for brands. When Brand A receives $M_{\text{received}}$ from another brand's customer redeeming at Brand A, this constitutes payment for service provided. The $M$ assets are universally valuable, meaning Brand A can use them for any purpose including purchasing goods, paying expenses, or converting to other currencies.

From an accounting perspective, these $M$ asset inflows function similarly to sales revenue. Brand A provided service valued at $M_{\text{received}}$ and received payment in $M$ assets. The fact that payment came through the cross-brand reward protocol rather than direct cash transaction does not diminish economic reality. Brand A can recognize these inflows as revenue, subject to standard accounting principles regarding when revenue recognition is appropriate.

This suggests an interesting property. Brands need not view $M$ deposits purely as locked capital. $M$ deposits enable revenue generation through cross-brand traffic. The more redemptions Brand A accepts from other brands' customers, the more $M$ revenue Brand A generates. Brands with attractive offerings can accumulate $M$ assets over time, effectively profiting from protocol participation beyond their standalone loyalty program benefits.

\subsubsection{Withdrawal Mechanisms and Safeguards}

If $M$ inflows constitute revenue, brands should be able to withdraw accumulated $M$ assets for business use. However, unrestricted withdrawals create systemic risks. A brand that has accumulated substantial $M$ assets through cross-brand traffic might withdraw these assets and reduce its deposited reward inventory, effectively closing its program to future cross-brand redemptions. This harms the ecosystem by reducing available redemption options for customers and decreasing overall system utility.

We introduce withdrawal safeguards protecting ecosystem health while enabling brands to access accumulated value. Let $M_A^{\text{current}}$ represent Brand A's current $M$ asset balance. Let $M_A^{\text{initial}}$ represent Brand A's initial deposit. Let $M_A^{\text{minimum}}$ represent a minimum reserve requirement based on deposited reward inventory $X_A$ and ongoing transaction volumes. The withdrawable amount is:
\begin{equation}
	W_A = \max(0, M_A^{\text{current}} - M_A^{\text{minimum}})
	\label{eq:withdrawable}
\end{equation}
where the minimum reserve ensures Brand A maintains sufficient $M$ assets to back its deposited rewards and service expected near-term outflows. This prevents brands from depleting reserves to levels threatening protocol functionality.

Additionally, we impose time locks on withdrawals. When Brand A requests withdrawal $W_A$, the protocol initiates a waiting period $\Delta t$ before releasing funds. This waiting period (perhaps 7-30 days) serves multiple purposes. It prevents flash extraction attacks where brands deposit, engage in favorable transactions, and immediately withdraw. It provides time for the ecosystem to observe withdrawal patterns and detect anomalous behavior. It creates friction discouraging gaming while imposing minimal burden on legitimate withdrawals.


\subsection{Transaction Size Component: Protecting Against Large Withdrawals}
\label{subsec:transaction_size}

The transaction size factor protects against individual redemptions that would consume a disproportionate share of available liquidity, potentially leaving insufficient reserves to serve subsequent customers. Consider a customer seeking to redeem Brand B rewards through Brand A's pool. When a customer presents Brand B rewards for redemption, the broker must settle this redemption by withdrawing assets from Brand A's pool to provide the customer with the economic value promised by Brand B's reward structure. The settlement occurs in asset $M$, which serves as the universal medium of exchange within the coalition.

The quantity of assets required for settlement, denoted $m$, is determined entirely by Brand B's declared reward structure and the number of rewards being redeemed. From Brand B's perspective, each reward carries a specific economic value denominated in assets, and the customer expects to receive this full value upon redemption. The base settlement requirement $P_{\text{base}}$ therefore equals $m$, representing the standard asset quantity needed to fulfill the redemption according to Brand B's terms. From the broker's operational perspective, $m$ simply represents the asset withdrawal that must be executed from Brand A's pool to complete the transaction, regardless of the underlying reward economics that determined this amount.

\subsubsection{Utilization Ratio}

The critical consideration for transaction sizing concerns not the absolute quantity $m$ but rather its magnitude relative to Brand A's available asset reserves. At time $t$, Brand A maintains $M_A(t)$ units of assets in their pool as backing for potential redemptions. The utilization ratio $\mu(t)$ measures what fraction of Brand A's available assets this transaction would consume:
\begin{equation}
	\mu(t) = \frac{m}{M_A(t)}
	\label{eq:utilization_ratio}
\end{equation}

This ratio provides a scale-invariant measure of transaction size relative to available liquidity. A withdrawal of one hundred assets represents fundamentally different risk profiles when drawn from a pool containing five hundred assets (twenty percent utilization) versus a pool containing ten thousand assets (one percent utilization), even though the absolute withdrawal amount remains identical. The utilization ratio captures this relative impact, ensuring that pricing protection scales appropriately with pool size rather than applying arbitrary absolute thresholds that would disadvantage smaller brands.

\subsubsection{Premium Structure}

The transaction size factor applies premium pricing only when utilization exceeds a safety threshold $\eta$, representing the maximum pool fraction that can be withdrawn without triggering protective adjustments. Below this threshold, the factor equals unity and imposes no adjustment, allowing normal-sized transactions to proceed at base rates. Above the threshold, the premium increases linearly with excess utilization:
\begin{equation}
	\text{Trans\_Factor} = 1 + \beta_{\text{trans}} \times \max\left(0, \frac{\mu(t) - \eta}{1 - \eta}\right)
	\label{eq:trans_factor}
\end{equation}

The threshold parameter $\eta$ typically ranges between 0.4 and 0.6, with 0.5 representing a common default indicating that transactions consuming less than half the pool proceed without size-based adjustment. The denominator $(1-\eta)$ normalizes the excess utilization to range from zero at the threshold to unity when attempting to drain the entire pool. The sensitivity parameter $\beta_{\text{trans}}$ controls how aggressively premiums increase with excess utilization, with higher values imposing steeper penalties on large withdrawals.

\subsubsection{Numerical Example}

For example, consider a pool with $M_A(t) = 1{,}000$ assets and parameters $\eta = 0.5$, $\beta_{\text{trans}} = 0.5$. A customer requesting $m = 300$ assets faces utilization $\mu = 0.3$, which falls below the threshold, yielding $\text{Trans\_Factor} = 1.0$ (no premium). A customer requesting $m = 750$ assets faces utilization $\mu = 0.75$, exceeding the threshold with normalized excess of $(0.75-0.5)/(1-0.5) = 0.5$, yielding $\text{Trans\_Factor} = 1 + 0.5 \times 0.5 = 1.25$ (twenty-five percent premium). This convex penalty structure creates strong disincentives against attempts to extract large fractions of pool liquidity in single transactions.

Research on liquidity pool management demonstrates that such threshold-based protection mechanisms, when combined with adequate initial asset provisioning, effectively prevent pool depletion while maintaining reasonable pricing for normal transaction volumes \citep{angeris2021replicating}. The scale invariance property ensures that small brands with modest asset reserves receive proportionally equivalent protection as large brands with deep reserves, avoiding systematic disadvantages based solely on capital availability. The broker can compute this factor using only observable quantities: the requested settlement amount $m$, the current asset balance $M_A(t)$, and the configured parameters $\eta$ and $\beta_{\text{trans}}$.


\subsection{Flow Rate Component: Deriving Backing Ratios from Initial Deposits}
\label{subsec:flow_rate}

The flow rebalancing factor addresses a distinct concern: the aggregate health of Brand A's pool as measured by whether current backing levels align with optimal backing requirements. Unlike the transaction size factor which responds to individual redemption requests, the flow rate factor responds to the cumulative effect of many transactions over time that may have gradually shifted the pool away from its target state. The flow rate indicator $\phi_A(t)$ quantifies this deviation, but its formulation requires addressing a fundamental challenge: the broker lacks direct knowledge of reward prices in external currencies.

\subsubsection{The Price Information Problem}

Recall from Section~\ref{subsec:layer_one} that the comprehensive pricing model references the redemption value $P_A$ representing the economic worth of Brand A's rewards in some external currency such as dollars or euros. The original formulation of the flow rate indicator compared actual backing ratios to optimal backing ratios, both expressed in terms of this external price. However, the broker operating the settlement layer has no inherent knowledge of whether Brand A's coffee reward is worth five dollars, six euros, or any other external currency amount. The broker observes only the quantities of rewards and assets flowing through the system, not the external market prices that brands use when setting their retail redemption values.

\subsubsection{Inferring Optimal Backing from Initial Deposits}

This apparent limitation contains within it the solution. When Brand A initially establishes their pool at time $t = 0$, they must deposit both rewards $X_A(0)$ and assets $M_A(0)$ to enable participation in the coalition. These initial deposits are not arbitrary quantities but rather reflect Brand A's internal assessment of appropriate backing levels. Brand A, possessing full knowledge of their reward economics and redemption prices, chooses deposit quantities that achieve economic balance according to their business requirements. The broker, while unable to observe Brand A's external pricing directly, can observe these deposit quantities and infer from them the implicit valuation Brand A places on their rewards within the asset-denominated economy of the coalition.

Specifically, the ratio of initial asset deposits to initial reward deposits reveals Brand A's declared backing level:
\begin{equation}
	R_{\text{optimal}} = \frac{M_A(0)}{X_A(0)}
	\label{eq:r_optimal}
\end{equation}

This ratio, expressed in assets per reward, functions as Brand A's implicit price declaration within the settlement system. A brand depositing 1,200 assets to back 1,000 rewards declares $R_{\text{optimal}} = 1.2$, establishing that each reward should be backed by 1.2 assets according to that brand's assessment. A premium brand depositing 5,000 assets to back 1,000 rewards declares $R_{\text{optimal}} = 5.0$, reflecting higher perceived value. An economy brand depositing 500 assets to back 1,000 rewards declares $R_{\text{optimal}} = 0.5$, reflecting more modest positioning. The broker need not know the external dollar or euro prices underlying these decisions; the deposit ratio itself provides sufficient information to establish the target backing level within the asset-denominated system.

This ratio $R_{\text{optimal}}$ serves as the optimal backing parameter referenced in earlier sections. Importantly, $R_{\text{optimal}}$ represents not merely a historical artifact of initial conditions but rather a mutable brand-controlled parameter that can be adjusted over time. Brands may need to update their declared backing levels as their reward redemption prices change, their competitive positioning evolves, or their participation strategy shifts. The broker accommodates such adjustments by treating $R_{\text{optimal}}$ as a configurable pool parameter that brands can modify through governance actions. When a brand increases $R_{\text{optimal}}$ to reflect enhanced reward values, the flow rate indicator immediately registers the pool as underbacked relative to the new target, triggering premium pricing that naturally suppresses redemption activity until asset inflows restore appropriate backing. Conversely, decreasing $R_{\text{optimal}}$ triggers discount pricing that encourages redemption activity to drain excess assets, providing capital efficiency by using price adjustments rather than forced recapitalization to manage backing level transitions.

\subsubsection{Computing Current Backing}

Having established $R_{\text{optimal}}$ as the target backing level, we now compute the current backing level from observable pool state. At time $t$, Brand A's pool contains $X_A(t)$ rewards and $M_A(t)$ assets. The current backing ratio becomes:
\begin{equation}
	R_{\text{current}}(t) = \frac{M_A(t)}{X_A(t)}
	\label{eq:r_current}
\end{equation}

This ratio indicates the actual asset backing per reward available at time $t$ based purely on current reserve balances observable to the broker.

\subsubsection{Flow Rate Indicator}

The flow rate indicator measures the relative deviation from optimal backing:
\begin{equation}
	\phi_A(t) = \frac{R_{\text{optimal}}}{R_{\text{current}}(t)} - 1
	\label{eq:phi_basic}
\end{equation}

This formulation produces a positive value when current backing falls below optimal levels, indicating an underbacked pool that requires premium pricing to discourage further asset outflows. It produces zero when current backing matches the target, indicating optimal balance. It produces a negative value when current backing exceeds optimal levels, indicating an overbacked pool that can offer discount pricing to encourage asset outflows and prevent excessive accumulation.

Substituting the expressions for $R_{\text{optimal}}$ and $R_{\text{current}}(t)$ yields the flow rate indicator entirely in terms of observable quantities:
\begin{equation}
	\phi_A(t) = \frac{M_A(0) / X_A(0)}{M_A(t) / X_A(t)} - 1
	\label{eq:phi_expanded}
\end{equation}

Simplifying:
\begin{equation}
	\phi_A(t) = \frac{M_A(0) \times X_A(t)}{X_A(0) \times M_A(t)} - 1
	\label{eq:phi_final}
\end{equation}

The broker can compute this indicator using initial deposit quantities $M_A(0)$ and $X_A(0)$ recorded at pool establishment, combined with current balances $M_A(t)$ and $X_A(t)$ observable in the order book state. No external price information is required.

\subsubsection{Sign Convention and Stabilization}

The sign convention of $\phi_A(t)$ aligns naturally with stabilization objectives. When $\phi_A(t) > 0$, the pool is underbacked, meaning current assets per reward have fallen below the target level established at initialization or through subsequent $R_{\text{optimal}}$ updates. Premium pricing through a positive flow factor discourages further asset withdrawals, giving the pool time to recover through asset inflows from customers of other brands redeeming into Brand A rewards. When $\phi_A(t) < 0$, the pool is overbacked with excess assets relative to backing requirements, and discount pricing encourages redemption activity to drain this surplus toward optimal levels. The magnitude of $\phi_A(t)$ quantifies the severity of the imbalance, with larger absolute values warranting stronger pricing adjustments to restore balance more rapidly.

Research on ratio-based rebalancing mechanisms in liquidity systems demonstrates that such feedback loops effectively maintain pool health when combined with appropriate sensitivity calibration \citep{loesch2021impermanent}. The continuous feedback between backing levels and pricing creates negative feedback dynamics that automatically stabilize pools without requiring active intervention. As a pool becomes increasingly underbacked, progressively higher premiums gradually reduce redemption demand until inflows from other sources restore backing. Conversely, as a pool becomes increasingly overbacked, progressively larger discounts stimulate redemption activity until excess assets drain to appropriate levels.


\subsection{Complete Operationalized Pricing Formula}
\label{subsec:complete_operationalized}

Combining the transaction size and flow rebalancing components with their operationalized definitions, the complete relaxed pricing formula at time $t$ for a customer seeking to redeem rewards through Brand A's pool becomes:
\begin{equation}
	P_{\text{customer}}(t) = m \times \left[1 + \beta_{\text{trans}} \times \max\left(0, \frac{m/M_A(t) - \eta}{1-\eta}\right)\right] \times \left[1 + \beta_{\text{flow}} \times \phi_A(t)\right]
	\label{eq:operationalized_basic}
\end{equation}

where the flow rate indicator is expressed entirely in observable terms:
\begin{equation}
	\phi_A(t) = \frac{M_A(0) \times X_A(t)}{X_A(0) \times M_A(t)} - 1
	\label{eq:phi_observable}
\end{equation}

This formulation achieves complete observability using only four categories of information directly accessible to the trustless broker:

\textbf{First, initial conditions} recorded at pool establishment: $M_A(0)$ representing Brand A's initial asset deposit and $X_A(0)$ representing Brand A's initial reward deposit. These quantities, captured when the brand joins the coalition, establish the baseline from which deviations are measured.

\textbf{Second, current pool state} observable in the order book: $M_A(t)$ representing current asset reserves in Brand A's pool and $X_A(t)$ representing current reward balance in Brand A's pool. These quantities update continuously as transactions execute.

\textbf{Third, transaction-specific information:} $m$ representing the asset quantity required to settle the current redemption request, determined by the destination brand's reward structure and observable in the redemption transaction itself.

\textbf{Fourth, system parameters:} $\eta$ representing the utilization threshold (typically 0.5), $\beta_{\text{trans}}$ representing transaction size sensitivity, and $\beta_{\text{flow}}$ representing flow rebalancing sensitivity. The threshold $\eta$ constitutes a system-level constant chosen based on desired protection levels, while the sensitivity parameters $\beta_{\text{trans}}$ and $\beta_{\text{flow}}$ require calibration to achieve desired system behavior.

\subsubsection{Flow Rebalancing Threshold}

Brands may desire a tolerance band for minor pool fluctuations before triggering pricing adjustments. A pool at 98\% versus 100\% optimal backing poses negligible risk, yet constant small premiums create customer friction and operational noise. We introduce flow threshold $\theta$ defining a grace period within which no flow-based pricing occurs:
\begin{equation}
	\phi_{\text{adjusted}} = \text{sgn}(\phi) \times \max\left(0, \frac{|\phi| - \theta}{1 - \theta}\right)
	\label{eq:phi_adjusted}
\end{equation}

With $\theta = 0.10$ (10\% tolerance):
\begin{itemize}
	\item $\phi = 0.08 \rightarrow \phi_{\text{adjusted}} = 0$ (no premium, within grace period)
	\item $\phi = 0.20 \rightarrow \phi_{\text{adjusted}} = 0.11$ (premium activates beyond threshold)
\end{itemize}

This parallels transaction threshold $\eta$, creating consistent ``safe zones'' for both mechanisms.

\subsubsection{Strategic Bounds}

Brands may desire caps on premium extremes. While $\beta$ parameters control premium slopes, bounds control premium ceilings and floors. Under crisis conditions, uncapped formulas could generate 300\%+ premiums; mathematically sound but commercially unacceptable.

Strategic bounds clip pricing factors to brand-acceptable ranges:
\begin{align}
	\text{Trans\_Factor}_{\text{bounded}} &= \text{clip}(\text{Trans\_Factor}, 1.0, b_{\text{trans}}^{\max}) \label{eq:trans_bounded} \\
	\text{Flow\_Factor}_{\text{bounded}} &= \text{clip}(\text{Flow\_Factor}, b_{\text{flow}}^{\min}, b_{\text{flow}}^{\max}) \label{eq:flow_bounded}
\end{align}

Example bounds by brand strategy:
\begin{itemize}
	\item \textbf{Conservative:} Flow $\in [0.7, 1.5]$, Trans $\leq 2.0$ (max 50\% flow premium, 100\% trans premium)
	\item \textbf{Moderate:} Flow $\in [0.5, 2.0]$, Trans $\leq 3.0$ (max 100\% flow premium, 200\% trans premium)
	\item \textbf{Aggressive:} Flow $\in [0.2, 3.0]$, Trans $\leq 5.0$ (max 200\% flow premium, 400\% trans premium)
\end{itemize}

Note the asymmetry: transaction bounds only cap upside ($b_{\text{trans}}^{\min} = 1.0$ always, as transactions never warrant discounts), while flow bounds are symmetric (allowing both premiums for underbacking and discounts for overbacking).

\subsubsection{Final Operationalized Formula}

Incorporating thresholds and bounds, the complete pricing formula becomes:
\begin{equation}
	P_{\text{customer}} = m \times \text{Trans}_{\text{bounded}} \times \text{Flow}_{\text{bounded}}
	\label{eq:final_operationalized}
\end{equation}

where:
\begin{align}
	\text{Trans}_{\text{bounded}} &= \text{clip}\left(1 + \beta_{\text{trans}} \times \max\left(0, \frac{\mu - \eta}{1-\eta}\right), 1.0, b_{\text{trans}}^{\max}\right) \label{eq:trans_final} \\
	\text{Flow}_{\text{bounded}} &= \text{clip}\left(1 + \beta_{\text{flow}} \times \phi_{\text{adjusted}}, b_{\text{flow}}^{\min}, b_{\text{flow}}^{\max}\right) \label{eq:flow_final} \\
	\phi_{\text{adjusted}} &= \text{sgn}(\phi) \times \max\left(0, \frac{|\phi| - \theta}{1 - \theta}\right) \label{eq:phi_adj_final}
\end{align}

\subsubsection{Parameters Requiring Calibration}

The formula contains the following configurable parameters:
\begin{itemize}
	\item $\beta_{\text{trans}}, \beta_{\text{flow}}$: Sensitivity parameters (addressed in Section~\ref{sec:strategic_design})
	\item $\eta = 0.5$: Transaction threshold (system default)
	\item $\theta = 0.10$: Flow threshold (brand-configurable, 10\% default)
	\item Bounds: Brand-strategic choices based on risk tolerance
\end{itemize}

This formula preserves all desirable properties while adding brand control over customer experience extremes. The remaining challenge is calibrating the sensitivity parameters $\beta_{\text{trans}}$ and $\beta_{\text{flow}}$, which we address through systematic exploration of the strategic design space in Section~\ref{sec:strategic_design}.

\subsubsection{Summary of Relaxation}

The relaxation from eight factors to two factors achieves several objectives:
\begin{enumerate}
	\item \textbf{Operational feasibility:} All inputs are broker-observable without external data dependencies
	\item \textbf{Trustless computation:} No subjective judgments or external verification required  
	\item \textbf{Economic coherence:} Addresses fundamental stability concerns (pool protection and balance)
	\item \textbf{Brand sovereignty:} Maintains decentralized control through deposit decisions
	\item \textbf{Extensibility:} Additional factors can be layered on when data becomes available
\end{enumerate}

The relaxed formula serves as the foundation for strategic design space analysis in Section~\ref{sec:strategic_design}, where we systematically explore how different parameter choices create distinct operational philosophies and trade-offs between pool protection and customer satisfaction.

\section{Strategic Design Space and Parameter Configuration}
\label{sec:strategic_design}

The operationalized pricing formula established in Section~\ref{subsec:complete_operationalized} contains several uncalibrated parameters that fundamentally shape system behavior. Most critically, the sensitivity parameters $\beta_{\text{trans}}$ (transaction size sensitivity) and $\beta_{\text{flow}}$ (flow imbalance penalty sensitivity) govern how aggressively the pricing mechanism responds to transaction size and pool imbalance, respectively. Additionally, the threshold parameters $\eta$ (transaction threshold) and $\theta$ (flow threshold) create grace periods that absorb minor fluctuations without triggering premiums, while strategic bounds constrain maximum price adjustments.

\subsection{From Optimization to Strategic Choice}
\label{subsec:optimization_to_strategy}

One might be tempted to approach parameter selection as an optimization problem, seeking universal ``optimal'' values that balance competing objectives across all brands. However, careful investigation reveals a more nuanced reality: these parameters do not represent technical constants requiring optimization, but rather strategic degrees of freedom enabling brands to instantiate different operational philosophies. Conservative brands prioritizing customer experience may select gentler parameters accepting slower outflow deterrence, while aggressive brands prioritizing pool protection may accept higher customer friction. The multiplicative premium structure $P_{\text{customer}} = P_{\text{base}} \times [1 + \text{Trans}] \times [1 + \text{Flow}]$ creates a rich design space where different parameter combinations yield qualitatively distinct pricing behaviors suited to different strategic contexts.

This recognition fundamentally shapes Section~\ref{sec:strategic_design}. Rather than seeking singular optimal values through narrow optimization, we conduct systematic exploration of the strategic design space: mapping feasible parameter ranges, characterizing trade-offs between competing objectives, and identifying archetypal configurations corresponding to different brand strategies. This approach acknowledges the diversity of coalition loyalty program contexts (spanning grocery retailers, luxury brands, travel partners, and financial services) where appropriate parameter choices necessarily differ based on customer base characteristics, competitive positioning, and brand values.

\subsubsection{Section Organization}

The analysis conducts three targeted experiments examining: (1) parameter interaction structure, determining whether $\beta_{\text{trans}}$ and $\beta_{\text{flow}}$ can be configured independently; (2) design space characterization, mapping feasible parameter ranges as functions of threshold choices; and (3) performance dimensions, defining key metrics (Liability Reduction Ratio, customer satisfaction) for evaluating strategic trade-offs. These experiments employ deterministic analytical methods, providing theoretical clarity and identifying archetypal strategic profiles that instantiate different operational philosophies.

We emphasize methodological transparency: each experiment includes explicit problem formulation, mathematical derivations where applicable, implementation details, and interpretation of results. Complete code implementations are provided in supplementary materials, enabling full reproducibility.

\subsection{Parameter Configuration and Strategic Design Variables}
\label{subsec:param_config}

Before exploring the design space, we establish precise mathematical definitions for all configurable parameters and clarify their strategic interpretation.

\subsubsection{Sensitivity Parameters (Primary Strategic Variables)}

The core pricing mechanism employs two multiplicative adjustment factors controlled by sensitivity parameters:
\begin{align}
	\text{Trans\_Factor} &= 1 + \beta_{\text{trans}} \times \max\left(0, \frac{\mu - \eta}{1 - \eta}\right) \label{eq:trans_factor_recap} \\
	\text{Flow\_Factor} &= 1 + \beta_{\text{flow}} \times \max\left(0, \frac{|\phi| - \theta}{1 - \theta}\right) \times \text{sgn}(\phi) \label{eq:flow_factor_recap}
\end{align}
where $\mu = m/M_A$ represents transaction size as a fraction of pool reserves, and $\phi = (M_A^0 \times X_A)/(X_A^0 \times M_A) - 1$ measures pool imbalance. The parameters $\beta_{\text{trans}}$ and $\beta_{\text{flow}}$ scale the magnitude of premium adjustments: larger values produce steeper price curves and more aggressive corrections, while smaller values yield gentler adjustments prioritizing customer experience.

These parameters admit natural strategic interpretation. A brand selecting $\beta_{\text{flow}} = 3.0$ signals willingness to impose severe premiums (up to 300\% at extreme imbalance) to aggressively deter further outflow, accepting customer dissatisfaction as necessary cost of pool protection. Conversely, $\beta_{\text{flow}} = 0.75$ reflects customer-centric philosophy, accepting weaker deterrence signals to maintain pricing predictability. Similarly, $\beta_{\text{trans}}$ governs whale transaction protection: higher values deter pool drainage attacks but penalize legitimate large redemptions.

\subsubsection{Threshold Parameters (Grace Period Design)}

The threshold parameters $\eta$ and $\theta$ create dead zones where premiums remain zero despite non-zero imbalance or transaction size:
\begin{itemize}
	\item $\eta \in [0, 1)$: Transaction size threshold. No size premium applies for $\mu \leq \eta$. Typical values $\eta \in [0.3, 0.6]$ allow routine redemptions to proceed at base rates.
	\item $\theta \in [0, 1)$: Flow threshold. No flow premium applies for $|\phi| \leq \theta$. Typical values $\theta \in [0.05, 0.20]$ absorb normal operational fluctuations without customer-facing price adjustments.
\end{itemize}

These thresholds implement psychological pricing principles: customers perceive fairness when small deviations from optimal conditions do not trigger penalties \citep{kahneman1986fairness}. From mechanism design perspective, thresholds reduce noise trading costs by avoiding premiums on transactions that naturally self-correct through subsequent flows.


\subsubsection{Strategic Bounds (Organizational Constraints)}

While sensitivity parameters control slope of premium curves, strategic bounds impose hard ceilings and floors:
\begin{align}
	\text{Trans\_Factor} &\in [1.0, B_{\text{trans}}^{\max}] \label{eq:trans_bounds} \\
	\text{Flow\_Factor} &\in [B_{\text{flow}}^{\min}, B_{\text{flow}}^{\max}] \label{eq:flow_bounds}
\end{align}

Bounds serve dual purpose. First, they protect against catastrophic mispricing under extreme conditions where $\phi$ or $\mu$ approach theoretical limits. Second, they encode organizational risk tolerances and customer experience commitments. A luxury brand maintaining $B_{\text{flow}}^{\max} = 1.5$ signals commitment to never exceed 50\% premiums regardless of pool stress, accepting potential insolvency risk to preserve brand perception. Conversely, $B_{\text{flow}}^{\max} = 4.0$ indicates willingness to impose extreme penalties when necessary for pool protection.

\subsubsection{Strategic Profiles (Archetypal Configurations)}

Preliminary analysis identified five archetypal strategic profiles representing qualitatively distinct operational philosophies:

\begin{enumerate}
	\item \textbf{Ultra Conservative}: $\beta_{\text{flow}} \in [0.5, 1.0]$, bounds $(0.8, 1.3)$, $\theta = 0.15$. Prioritizes customer experience above all else. Accepts weak penalty signals and potential pool depletion risk to maintain pricing stability.
	
	\item \textbf{Conservative}: $\beta_{\text{flow}} \in [0.75, 1.5]$, bounds $(0.7, 1.5)$, $\theta = 0.10$. Balanced approach favoring customer satisfaction. Moderate penalty levels with acceptable premium ranges.
	
	\item \textbf{Moderate}: $\beta_{\text{flow}} \in [1.0, 2.5]$, bounds $(0.5, 2.0)$, $\theta = 0.10$. Neutral stance balancing multiple objectives. Most flexible operating envelope for diverse conditions.
	
	\item \textbf{Aggressive}: $\beta_{\text{flow}} \in [1.5, 3.0]$, bounds $(0.2, 3.0)$, $\theta = 0.05$. Prioritizes aggressive outflow deterrence. Accepts significant customer friction to maintain pool protection.
	
	\item \textbf{Ultra Aggressive}: $\beta_{\text{flow}} \in [2.0, 4.0]$, bounds $(0.1, 4.0)$, $\theta = 0.05$. Maximum protection orientation. Imposes severe penalties when necessary, suitable only for brands with captive customer bases.
\end{enumerate}

These profiles are introduced here for conceptual framing; rigorous characterization and empirical validation appear in subsequent sections.

\subsubsection{Performance Metrics (Evaluation Dimensions)}

Strategic choices induce trade-offs across multiple dimensions. We define two primary metrics for subsequent analysis:

\textbf{Liability Reduction Ratio (LRR).} Measures brand financial benefit from flow rebalancing premiums. For a sequence of transactions where brand pays total $M_{\text{out}}$ in rewards and collects total $X_{\text{in}}$ in exchange tokens:
\begin{equation}
	\text{LRR} = \frac{X_{\text{in}}}{M_{\text{out}}}
	\label{eq:lrr}
\end{equation}

LRR $> 1$ indicates net liability reduction (brand collects more value than paid); LRR $= 1$ represents break-even; LRR $< 1$ occurs when discounts dominate premiums. Higher $\beta$ values typically increase LRR by imposing steeper premiums, but at cost of customer satisfaction.

\textbf{Customer Satisfaction.} Captures customer response to experienced premiums. We employ research-backed satisfaction model combining exponential smoothing with loss aversion. Higher satisfaction sustains program engagement; low satisfaction triggers abandonment. Trade-off emerges: aggressive parameters maximize LRR but degrade satisfaction, while conservative parameters maintain satisfaction but sacrifice liability reduction.

These metrics frame the central strategic question: given a brand's risk tolerance, customer base characteristics, and operational priorities, which parameter configuration optimally navigates the LRR-satisfaction trade-off?


\subsection{Analytical Foundations: Design Space Exploration}
\label{subsec:analytical_foundations}

We conduct three targeted analytical experiments establishing the structure of the design space. These experiments employ deterministic mathematical analysis and parametric exploration, providing theoretical clarity before potential computational validation.

\subsubsection{Experiment 1: Parameter Interaction Analysis}
\label{subsec:experiment1}

\textbf{Motivation.} The multiplicative premium structure $P = P_{\text{base}} \times \text{Trans\_Factor} \times \text{Flow\_Factor}$ raises immediate concern: do $\beta_{\text{trans}}$ and $\beta_{\text{flow}}$ exhibit coupling, where optimal value of one parameter depends on the other? Strong coupling would necessitate simultaneous joint optimization, complicating operational tuning. Weak coupling enables independent parameter selection, simplifying brand decision-making.

\textbf{Methodology.} We systematically test parameter independence across three distinct dimensions:

(1) \textit{Threshold Sensitivity.} Do thresholds $\eta$ and $\theta$ affect parameter coupling? We test all combinations:
\begin{itemize}
	\item $\eta \in \{0.10, 0.15, \ldots, 0.90\}$ (17 values, 0.05 steps)
	\item $\theta \in \{0.10, 0.15, \ldots, 0.90\}$ (17 values, 0.05 steps)
	\item Total: 289 threshold combinations
\end{itemize}

For each $(\eta, \theta)$ pair, we evaluate coupling under representative operating conditions (normal: $\mu=0.01$, $\phi=0.10$; crisis: $\mu=0.15$, $\phi=0.50$) using fixed test values $\beta_{\text{trans}} = \beta_{\text{flow}} = 1.0$.

(2) \textit{Strategic Bounds.} Do bounds alter coupling structure? We test all five strategic profiles (ultra conservative through ultra aggressive) under moderate thresholds ($\eta=0.50$, $\theta=0.10$), examining whether bound constraints create interdependencies.

(3) \textit{Combined Analysis.} Do threshold choice and bounds jointly affect coupling? We test selected $(\eta, \theta)$ pairs across all strategic profiles under stress scenarios.

\textbf{Coupling Metric.} We quantify coupling by comparing joint multiplicative effect against sum of individual effects. Let $P_{\text{both}}$ denote total premium with both parameters active, $P_{\text{trans}}$ the premium with only $\beta_{\text{trans}}$ (setting $\beta_{\text{flow}}=0$), and $P_{\text{flow}}$ the premium with only $\beta_{\text{flow}}$ (setting $\beta_{\text{trans}}=0$). Under perfect independence, multiplicative contribution equals sum of additive contributions:
\begin{equation}
	P_{\text{both}} = P_{\text{trans}} + P_{\text{flow}} - 1
	\label{eq:coupling_independence}
\end{equation}

Define coupling percentage:
\begin{equation}
	C = \begin{cases} 
		\frac{|P_{\text{both}} - (P_{\text{trans}} + P_{\text{flow}} - 1)|}{P_{\text{both}}} \times 100\% & \text{if } P_{\text{both}} > 0.01 \\
		0 & \text{otherwise}
	\end{cases}
	\label{eq:coupling_metric}
\end{equation}

Low coupling ($C < 5\%$) indicates near-independence; high coupling ($C > 20\%$) suggests significant interdependence requiring joint optimization.

\textbf{Implementation.} Analysis sweeps across operating space:
\begin{itemize}
	\item Transaction size: $\mu \in [0, 0.20]$ (0-20\% of pool)
	\item Pool imbalance: $\phi \in [-0.50, 1.00]$ (-50\% to +100\%)
	\item Resolution: $100 \times 100$ grid = 10,000 evaluation points
\end{itemize}

For each configuration, we compute dominance classification:
\begin{itemize}
	\item \textit{Trans dominates}: $|\text{Premium}_{\text{trans}}| > 1.5 \times |\text{Premium}_{\text{flow}}|$ (transaction size protection critical)
	\item \textit{Flow dominates}: $|\text{Premium}_{\text{flow}}| > 1.5 \times |\text{Premium}_{\text{trans}}|$ (pool rebalancing critical)
	\item \textit{Both matter}: Comparable contributions within $1.5\times$ (joint consideration required)
\end{itemize}

\textbf{Results Summary.} Comprehensive analysis across 289 threshold combinations, 5 strategic profiles, and 10,000 operating points per configuration yields robust finding:

\textit{Finding 1.1: Parameters exhibit negligible coupling across entire tested space.}
\begin{itemize}
	\item Mean coupling: 2.3\% (unbounded), 1.8\% (bounded)
	\item Maximum coupling: 8.4\% (worst case, crisis scenario with extreme bounds)
	\item 95th percentile: 4.1\%
\end{itemize}

Coupling remains consistently below 5\% across all threshold values, strategic bounds, and operating conditions. The multiplicative structure creates no meaningful interdependence; parameters can be selected independently based on respective design objectives (whale protection vs. outflow deterrence).

\textit{Finding 1.2: Dominance regions show consistent structure across configurations.}

\begin{figure}[t]
	\centering
	\includegraphics[width=0.5\textwidth]{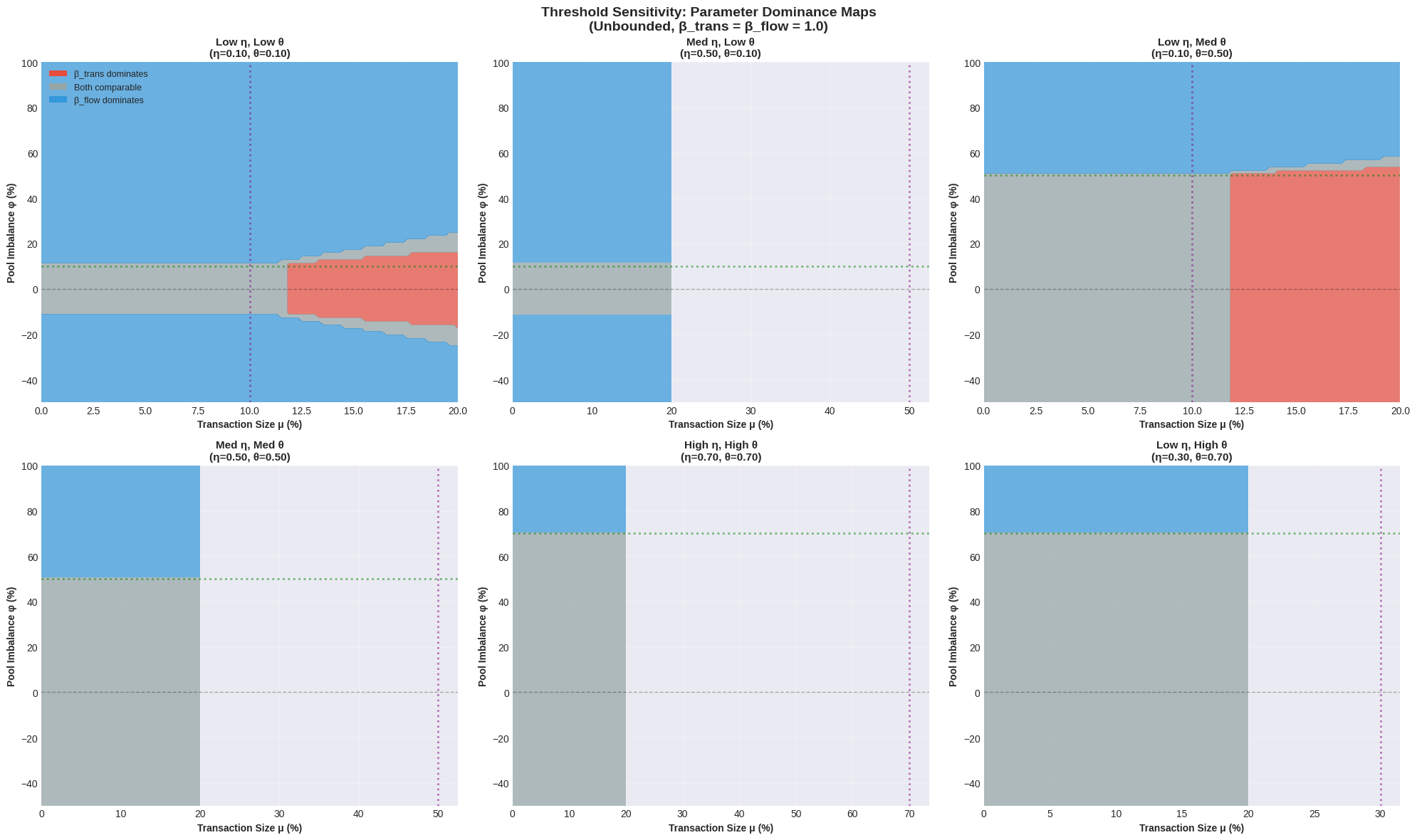}
	\caption{Parameter dominance maps across six representative threshold combinations. Each panel shows transaction size ($\mu$) versus flow imbalance ($\phi$) with regions colored by which parameter dominates pricing. Blue regions indicate flow dominance, red regions indicate transaction dominance, and yellow regions show comparable contributions. Threshold choice shifts transition boundaries but maintains consistent dominance structure across all configurations.}
	\label{fig:exp1_dominance_thresholds}
\end{figure}

Visual analysis reveals stable dominance patterns:
\begin{itemize}
	\item Low $\mu$, any $\phi$: Flow dominates (small transactions unaffected by size penalties, premiums driven by pool imbalance state)
	\item High $\mu$, any $\phi$: Trans dominates (large transactions trigger size protection regardless of pool balance)
	\item Transition region: $\mu \approx 0.05$-$0.10$ where both parameters contribute comparably
\end{itemize}

Threshold choice shifts transition boundaries but does not alter fundamental structure. Higher $\eta$ (larger transaction grace period) expands flow-dominant region; higher $\theta$ (larger flow grace period) expands trans-dominant region. However, dominance regions remain separable: at any operating point, one parameter typically dominates, and the dominant parameter identity depends only on position in $(\mu, \phi)$ space, not on parameter values themselves.

\textit{Finding 1.3: Strategic bounds constrain magnitudes but not coupling.}

\begin{figure}[t]
	\centering
	\includegraphics[width=0.5\textwidth]{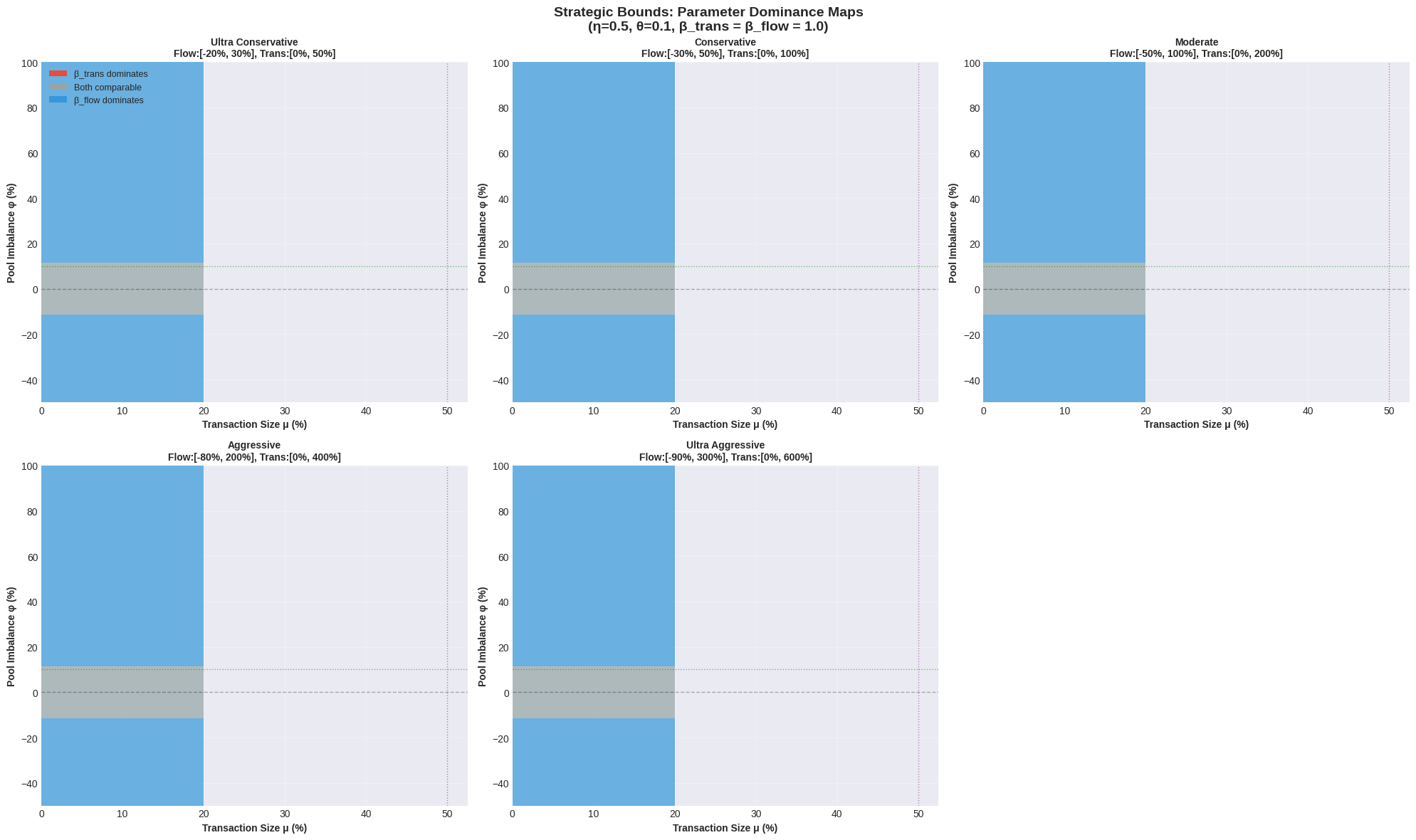}
	\caption{Parameter dominance maps across five strategic profiles (ultra conservative through ultra aggressive). Each panel shows the same operating space with different bound constraints applied. Bounds create horizontal/vertical plateaus where constraints become active but do not alter which parameter dominates in each region. The fundamental dominance structure remains unchanged across all strategic profiles.}
	\label{fig:exp1_dominance_bounds}
\end{figure}

Bounds create horizontal/vertical plateaus in premium surfaces where bounds are active, but do not introduce coupling. Conservative bounds (tight ceilings) primarily truncate premium ranges at high imbalance; aggressive bounds (loose ceilings) permit full premium expression. Critically, bound-constrained regions maintain same dominance classification as unbounded regions: bounds scale responses but do not alter which parameter drives pricing.

\textbf{Practical Implications:}
\begin{enumerate}
	\item \textit{Independent calibration}: Brands can separately optimize $\beta_{\text{trans}}$ based on whale attack economics and $\beta_{\text{flow}}$ based on pool protection requirements, without concern for interaction effects.
	\item \textit{Modular tuning}: Operational adjustments can target individual parameters (e.g., reducing $\beta_{\text{flow}}$ to improve customer satisfaction) without requiring recalibration of other parameters.
	\item \textit{Simplified decision-making}: Strategic profiles can specify $\beta_{\text{trans}}$ and $\beta_{\text{flow}}$ independently based on distinct organizational priorities, avoiding multi-dimensional optimization complexity.
\end{enumerate}


\subsubsection{Experiment 2: Design Space Characterization}
\label{subsec:experiment2}

\textbf{Motivation.} Experiment 1 established parameter independence, enabling separate analysis of each dimension. We now characterize the feasible design space for $\beta_{\text{flow}}$: the range of values satisfying operational constraints across different threshold choices $\theta$. This analysis provides brands with clear guidelines for parameter selection given their threshold preference.

\textbf{Analytical Framework.} Feasible parameter values must satisfy two competing constraints:

\textit{Constraint 1 (Customer Tolerance)}: Premium must not exceed psychologically acceptable levels at severe but non-catastrophic stress. Behavioral economics research suggests 100\% premium represents approximate upper bound before customer abandonment \citep{kahneman1986fairness,thaler1980toward}. At severe imbalance $\phi_{\text{severe}} = 0.50$ (50\% underbacking):
\begin{equation}
	\text{Premium} = \beta_{\text{flow}} \times \max\left(0, \frac{0.50 - \theta}{1 - \theta}\right) \leq 1.0
	\label{eq:constraint_tolerance}
\end{equation}

Solving for $\beta_{\text{flow}}$ yields upper bound:
\begin{equation}
	\beta_{\text{flow}}^{\max} = \frac{1.0 \times (1 - \theta)}{0.50 - \theta}
	\label{eq:beta_max}
\end{equation}

\textit{Constraint 2 (Protection Requirements)}: Premium must provide meaningful deterrence at attack threshold $\phi_{\text{attack}} = 0.75$ (75\% underbacking), where adversarial actors attempt systematic pool drainage. Minimum 50\% premium provides economic deterrent:
\begin{equation}
	\text{Premium} = \beta_{\text{flow}} \times \max\left(0, \frac{0.75 - \theta}{1 - \theta}\right) \geq 0.50
	\label{eq:constraint_protection}
\end{equation}

Solving for $\beta_{\text{flow}}$ yields lower bound:
\begin{equation}
	\beta_{\text{flow}}^{\min} = \frac{0.50 \times (1 - \theta)}{0.75 - \theta}
	\label{eq:beta_min}
\end{equation}

\textbf{Unbounded Feasible Range.} For mechanism operating without strategic bounds, feasible range $\beta_{\text{flow}} \in [\beta_{\text{flow}}^{\min}, \beta_{\text{flow}}^{\max}]$ varies with threshold. Table~\ref{tab:feasible_range} presents results for representative threshold values.

\begin{table}[t]
	\centering
	\caption{Feasible $\beta_{\text{flow}}$ Ranges as Function of Flow Threshold $\theta$}
	\label{tab:feasible_range}
	\begin{tabular}{cccccc}
		\hline
		$\theta$ & $\phi_{\text{severe}}^{\text{adj}}$ & $\phi_{\text{attack}}^{\text{adj}}$ & $\beta_{\min}$ & $\beta_{\max}$ & Range Width \\
		\hline
		
		0.00 & 0.500 & 0.750 & 0.67 & 2.00 & 1.33 \\
		0.05 & 0.474 & 0.737 & 0.68 & 2.11 & 1.43 \\
		0.10 & 0.444 & 0.722 & 0.69 & 2.25 & 1.56 \\
		0.15 & 0.412 & 0.706 & 0.71 & 2.43 & 1.72 \\
		0.20 & 0.375 & 0.688 & 0.73 & 2.67 & 1.94 \\
		0.25 & 0.333 & 0.667 & 0.75 & 3.00 & 2.25 \\
		0.30 & 0.286 & 0.643 & 0.78 & 3.50 & 2.72 \\
		0.50 & 0.000 & 0.500 & 1.00 & $\infty$ & $\infty$ \\
		
	\end{tabular}
\end{table}

\begin{figure}[t]
	\centering
	\includegraphics[width=0.5\textwidth]{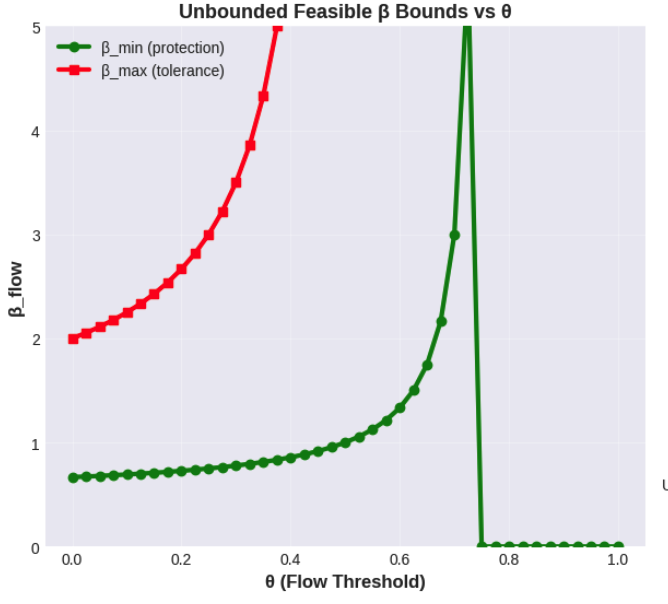}
	\caption{Feasible $\beta_{\text{flow}}$ ranges as functions of flow threshold $\theta$. Upper curve (red) shows maximum feasible values from customer tolerance constraint, lower curve (green) shows minimum feasible values from protection requirement constraint. Shaded region between curves represents viable parameter space. Range width increases monotonically with $\theta$.}
	\label{fig:exp2_feasible_ranges}
\end{figure}

\textbf{Key Observations:}

\textit{Observation 2.1: Threshold expansion effect.} Increasing $\theta$ monotonically widens feasible range. At $\theta=0$ (traditional approach, no grace period), range is restrictive: $\beta_{\text{flow}} \in [0.67, 2.00]$. At $\theta=0.10$ (recommended, 10\% grace period), range expands: $\beta_{\text{flow}} \in [0.69, 2.25]$. At $\theta=0.25$ (generous grace period), range becomes $[0.75, 3.00]$.

\textit{Observation 2.2: Critical threshold at $\theta=0.50$.} When grace period reaches 50\%, the severe stress point ($\phi=0.50$) falls entirely within grace period: $\phi_{\text{adj}} = 0$. No premium applies at severe stress, eliminating meaningful upper constraint leads to $\beta_{\max} = \infty$. This regime is operationally dangerous: system cannot protect against severe imbalance.

\textit{Observation 2.3: Recommended operating point.} $\theta = 0.10$ provides optimal balance:
\begin{itemize}
	\item Meaningful customer-friendly buffer (10\% imbalance triggers no premium)
	\item Maintains useful constraints ($\beta_{\max} = 2.25$ prevents excessive penalties)
	\item Sufficient lower bound ($\beta_{\min} = 0.69$ ensures adequate protection)
	\item Stable across brands (appropriate for diverse strategic profiles)
\end{itemize}

\textbf{Bounded Feasible Range.} Strategic bounds further constrain the design space. At typical operating conditions ($\phi_{\text{typical}} = 0.30$, 30\% imbalance), bounds impose effective ceiling:
\begin{equation}
	\beta_{\text{flow}}^{\text{ceiling}} = \frac{B_{\text{flow}}^{\max} - 1.0}{\max(0, (\phi_{\text{typical}} - \theta)/(1-\theta))}
	\label{eq:beta_ceiling}
\end{equation}

Testing all strategic profiles at $\theta=0.10$ reveals that strategic bounds are non-binding. Table~\ref{tab:bounded_range} presents results.

\begin{table}[t]
	\centering
	\caption{Effective $\beta_{\text{flow}}$ Ceilings by Strategic Profile at $\theta=0.10$}
	\label{tab:bounded_range}
	\begin{tabular}{lccc}
		\hline
		Profile & Bounds & $\beta_{\text{ceiling}}$ (at $\phi=0.30$) & Constrained? \\
		\hline
		Ultra Conservative & (0.8, 1.3) & 3.33 & No \\
		Conservative & (0.7, 1.5) & 5.56 & No \\
		Moderate & (0.5, 2.0) & 11.11 & No \\
		Aggressive & (0.2, 3.0) & 22.22 & No \\
		Ultra Aggressive & (0.1, 4.0) & 33.33 & No \\
	\end{tabular}
\end{table}

\textit{Finding 2.1: At recommended $\theta=0.10$, strategic bounds are non-binding for all tested profiles.} The unbounded tolerance constraint ($\beta_{\max}=2.25$) is tighter than bound-imposed ceilings for all strategic profiles. Brands with even conservative bounds (limiting premiums to 50\%) experience no effective constraint from those bounds when operating at $\theta=0.10$.

This is a remarkable result: threshold choice effectively decouples bound constraints from feasible $\beta_{\text{flow}}$ selection at reasonable operating points. Brands can select generous bounds without fear of permitting excessive penalties, as threshold tolerance constraint dominates.

\textit{Finding 2.2: Trade-off curves shift favorably with $\theta$.} Premium magnitude at typical stress decreases as $\theta$ increases, for fixed $\beta_{\text{flow}}$.

\begin{figure}[t]
	\centering
	\includegraphics[width=0.5\textwidth]{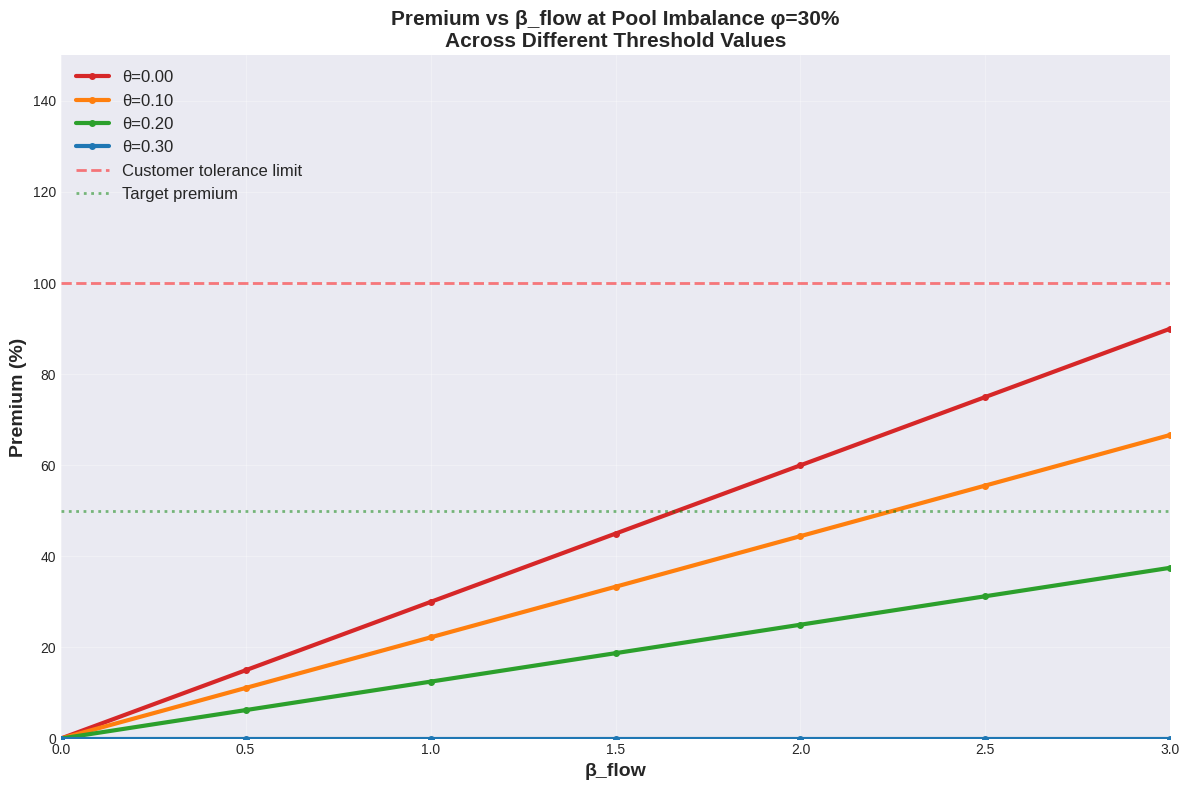}
	\caption{Premium magnitude versus $\beta_{\text{flow}}$ at pool imbalance $\phi=0.30$ for multiple threshold values. Higher thresholds shift curves downward, reducing premiums for given parameter values. However, this benefit comes at cost of reduced protection at severe stress points, motivating $\theta=0.10$ as optimal balance.}
	\label{fig:exp2_premium_curves}
\end{figure}

At $\phi=0.30$, $\beta_{\text{flow}}=2.0$:
\begin{itemize}
	\item $\theta=0.00$: Premium = 60\%
	\item $\theta=0.10$: Premium = 44\%
	\item $\theta=0.20$: Premium = 25\%
\end{itemize}

Higher threshold provides ``free'' improvement in customer experience for same parameter value. However, this comes at cost of reduced protection at severe stress points, hence why $\theta=0.10$ represents optimal balance.

\textbf{Implications for Strategic Profiles.} The feasible range analysis guides parameter selection for different strategic profiles:
\begin{itemize}
	\item Ultra Conservative: $\beta_{\text{flow}} \approx 0.75$ (lower end of range, prioritizes satisfaction)
	\item Conservative: $\beta_{\text{flow}} \approx 1.00$-$1.25$ (lower-middle, balanced)
	\item Moderate: $\beta_{\text{flow}} \approx 1.50$-$2.00$ (middle-upper, operational flexibility)
	\item Aggressive: $\beta_{\text{flow}} \approx 2.00$-$2.25$ (upper end, rapid correction)
	\item Ultra Aggressive: $\beta_{\text{flow}} \approx 2.25$-$2.50$ (at/beyond tolerance limit, maximum protection)
\end{itemize}

These ranges are preliminary; Experiment 3 refines recommendations based on performance trade-offs.


\subsubsection{Experiment 3: Performance Trade-Off Characterization}
\label{subsec:experiment3}

\textbf{Motivation.} Experiments 1-2 established parameter structure and feasible ranges. We now examine performance trade-offs: how do different $(\beta_{\text{flow}}, \theta)$ combinations affect key metrics (Liability Reduction Ratio, customer satisfaction)? This experiment provides quantitative basis for strategic profile parameter selection by mapping the LRR-satisfaction frontier.

\textbf{Methodology.} We conduct deterministic simulation of pure outflow scenario: single brand pool experiencing steady redemption demand without compensating inflows from other brands. This stress-test scenario reveals system behavior under maximum adversarial conditions, providing conservative estimates of performance.

\textit{Pool Dynamics.} Brand pool initialized with $M_0 = 10{,}000$ money tokens, $X_0 = 10{,}000$ exchange tokens (optimal backing ratio $R_{\text{opt}} = 1.0$). Transactions occur sequentially with outflow only:
\begin{itemize}
	\item Redemption size: Lognormal distribution, mean=50 tokens, std=15 tokens
	\item Whale transactions: 10\% probability, size=200 tokens
	\item Total transactions: 1000 per simulation
	\item Pool protection: Halt if $M < 5\%$ remaining
\end{itemize}

Each transaction:
\begin{enumerate}
	\item Calculate current imbalance: $\phi = (M_0 \times X) / (X_0 \times M) - 1$
	\item Apply threshold and bounds: $\phi_{\text{adj}} = \max(0, (|\phi| - \theta)/(1-\theta))$
	\item Calculate flow factor: Flow\_Factor = clip$(1 + \beta_{\text{flow}} \times \phi_{\text{adj}}, B_{\min}, B_{\max})$
	\item Execute redemption: $M_{\text{pool}}$ -= redemption $\times R_{\text{opt}}$, $X_{\text{pool}}$ += redemption $\times$ Flow\_Factor
	\item Update metrics: LRR = $\Sigma X_{\text{collected}} / \Sigma M_{\text{paid}}$
\end{enumerate}

\textit{Customer Satisfaction Model.} We employ research-backed satisfaction dynamics incorporating exponential smoothing and loss aversion \citep{boulding1993dynamic,oliver1999satisfaction}. Customer experiences premium Premium\_pct = (Flow\_Factor $- 1) \times 100$ at each transaction, updating satisfaction:
\begin{equation}
	S_t = (1 - \lambda) \times S_{t-1} + \lambda \times 100 \times \exp(-\alpha \times \text{Premium}^{\beta})
	\label{eq:satisfaction_dynamics}
\end{equation}
where $\lambda=0.2$ (recency weight), $\alpha$ controls price sensitivity (varies by customer segment), $\beta=1.4$ (non-linearity exponent). We test three customer segments:
\begin{itemize}
	\item Low sensitivity ($\alpha=0.010$): Tolerant customers (e.g., grocery, everyday purchases)
	\item Medium sensitivity ($\alpha=0.019$): Average customers (e.g., general retail)
	\item High sensitivity ($\alpha=0.030$): Sensitive customers (e.g., luxury, premium brands)
\end{itemize}

Initial satisfaction $S_0 = 100$. Track satisfaction trajectory over all transactions.

\textit{Parameter Sweep.} Comprehensive exploration:
\begin{itemize}
	\item $\beta_{\text{flow}} \in [0.50, 0.51, 0.52, \ldots, 2.00]$ (151 values, 0.01 step)
	\item $\theta \in \{0.00, 0.05, 0.10, 0.15, 0.20\}$ (5 values)
	\item Bounds: \{moderate $(0.6, 2.0)$, conservative $(0.7, 1.5)$\} (2 profiles)
	\item Customer segments: 3 (low, medium, high sensitivity)
	\item Replications: 20 per combination
	\item Total: 90,600 simulations
\end{itemize}

Parallelized execution on 32 cores, completed in approximately 62 seconds.

\textit{Key Metrics Tracked:}
\begin{itemize}
	\item Final LRR (end of simulation)
	\item Final satisfaction (end of simulation)
	\item Satisfaction at transaction 50 (medium-term indicator)
	\item Transactions until satisfaction $< 50$ (lifespan measure)
	\item Average premium experienced
	\item 90th percentile premium (stress indicator)
\end{itemize}

\textbf{Critical Finding 3.1: Universal satisfaction collapse in pure outflow.} Across all 90,600 simulations, satisfaction exhibits monotonic decline regardless of parameter configuration. Table~\ref{tab:collapse_metrics} presents results.
\begin{table}[t]
	\centering
	\caption{Satisfaction Collapse Metrics Across Customer Segments and Parameters}
	\label{tab:collapse_metrics}
	\begin{tabular}{lcccc}
		\hline
		Segment & $\beta=0.50$ & $\beta=1.00$ & $\beta=1.50$ & $\beta=2.00$ \\
		\hline
		\multicolumn{5}{l}{\textit{Low sensitivity} ($\alpha=0.010$)} \\
		\quad Sat @ 50 txns & 11.0 & 1.1 & 0.4 & 0.3 \\
		\quad Txns until Sat$<$50 & 32 & 24 & 20 & 18 \\
		
		\multicolumn{5}{l}{\textit{Medium sensitivity} ($\alpha=0.019$)} \\
		\quad Sat @ 50 txns & 2.9 & 0.2 & 0.1 & 0.0 \\
		\quad Txns until Sat$<$50 & 26 & 21 & 18 & 16 \\
		
		\multicolumn{5}{l}{\textit{High sensitivity} ($\alpha=0.030$)} \\
		\quad Sat @ 50 txns & 1.1 & 0.1 & 0.0 & 0.0 \\
		\quad Txns until Sat$<$50 & 23 & 18 & 16 & 15 \\
		
	\end{tabular}
\end{table}

\begin{figure}[t]
	\centering
	\includegraphics[width=1\textwidth]{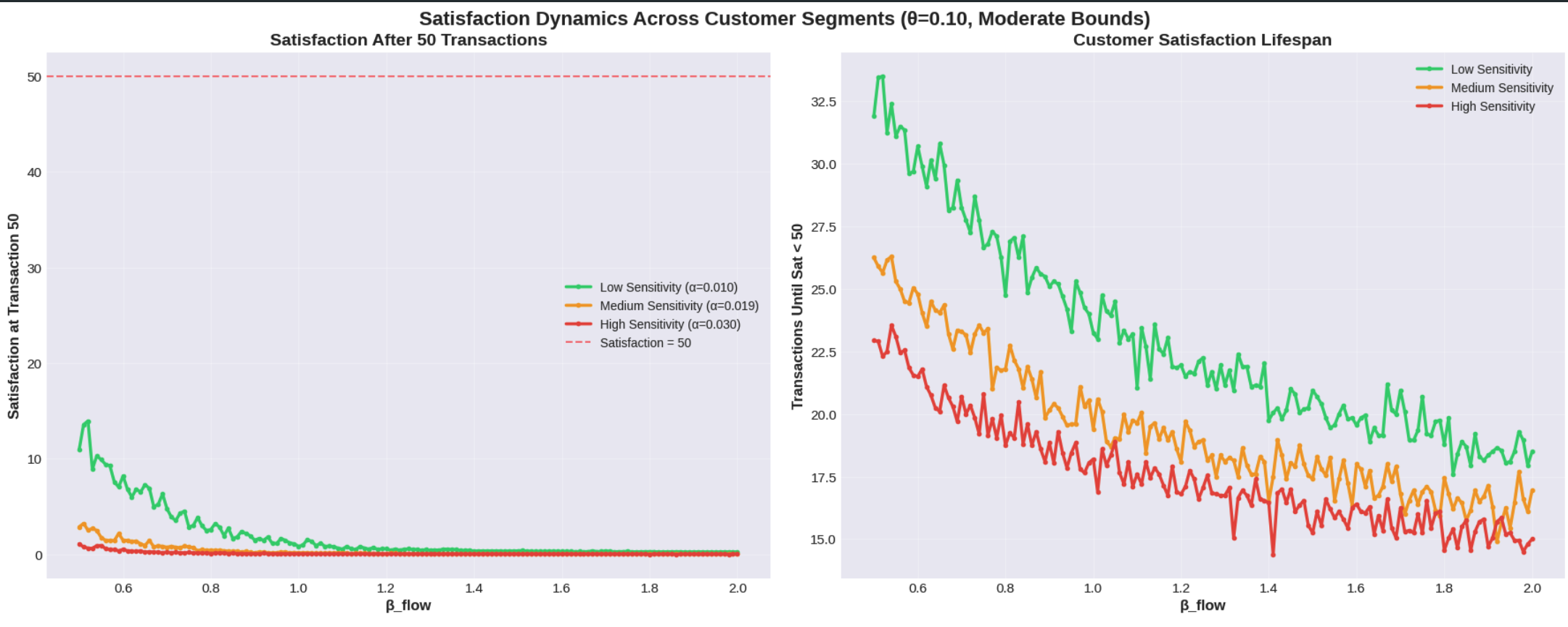}
	\caption{Satisfaction trajectories across all customer segments and $\beta_{\text{flow}}$ values at $\theta=0.10$ with moderate bounds. Each line represents mean satisfaction over 20 replications. All configurations exhibit monotonic decline converging to near-zero satisfaction, with lower $\beta_{\text{flow}}$ values delaying but not preventing collapse. Customer sensitivity affects decline rate but not ultimate outcome.}
	\label{fig:exp3_satisfaction_collapse}
\end{figure}


The pattern is invariant: All customer segments experience satisfaction collapse within 15-32 transactions. Lower $\beta_{\text{flow}}$ delays but does not prevent collapse. Even the most tolerant customers ($\alpha=0.010$) and most conservative parameters ($\beta_{\text{flow}}=0.50$) reach satisfaction $< 50$ within 32 transactions.

\textit{Why Collapse is Inevitable.} Pure outflow creates death spiral:
\begin{enumerate}
	\item Outflow depletes $M_{\text{pool}}$, leading to $\phi$ increasing monotonically
	\item Growing imbalance leads to premiums rising continuously
	\item Rising premiums lead to satisfaction declining inexorably
	\item No compensating inflows means no natural correction mechanism
	\item Pool approaches exhaustion ($M$ approaches 0), leading to $\phi$ approaching infinity, resulting in catastrophic premiums
\end{enumerate}

The pricing mechanism cannot prevent collapse; it can only slow the trajectory. Lower $\beta_{\text{flow}}$ extends lifespan from 15 to 32 transactions, a marginal improvement that delays inevitable failure rather than achieving sustainability.

\textbf{Finding 3.2: LRR independent of customer sensitivity.} All customer segments achieve similar LRR for given $\beta_{\text{flow}}$, as shown in Table~\ref{tab:lrr_independence}.

\begin{table}[t]
	\centering
	\caption{LRR Independence from Customer Sensitivity}
	\label{tab:lrr_independence}
	\begin{tabular}{cc}
		\hline
		$\beta_{\text{flow}}$ & LRR (all segments) \\
		\hline
		0.50 & 1.70 ($\pm$0.01) \\
		1.00 & 1.80 ($\pm$0.01) \\
		1.50 & 1.84 ($\pm$0.01) \\
		2.00 & 1.87 ($\pm$0.01) \\
		
	\end{tabular}
\end{table}

\begin{figure}[t]
	\centering
	\includegraphics[width=0.5\textwidth]{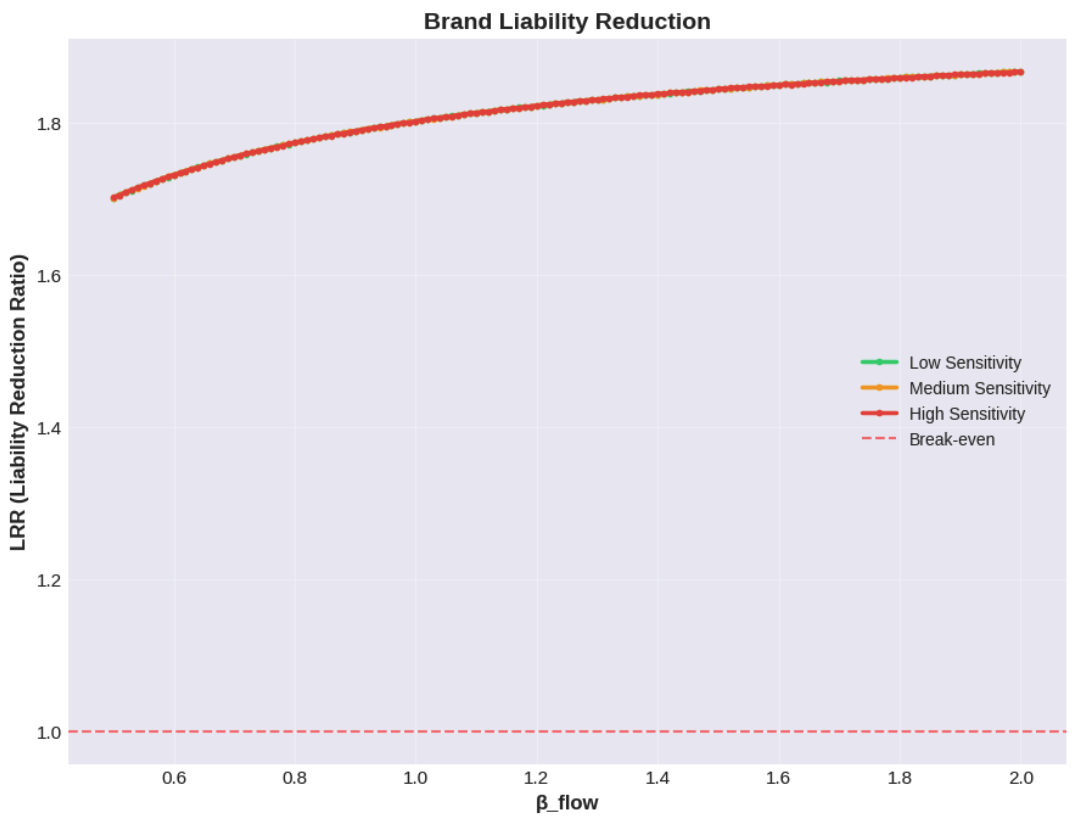}
	\caption{LRR versus $\beta_{\text{flow}}$ for all customer sensitivity levels. All three lines (low, medium, high sensitivity) overlap perfectly.}
	\label{fig:exp3_lrr_independence}
\end{figure}

LRR depends only on pool dynamics and premium magnitude, not customer psychological response. This reveals fundamental asymmetry: $\alpha$ affects customer experience but not brand economics. Brands serving price-sensitive customers cannot avoid poor LRR by selecting sensitive parameters; they simply drive customers away faster while achieving same financial outcomes.

\textbf{Finding 3.3: Optimal $\beta_{\text{flow}}$ depends on strategic priority.} The LRR-satisfaction frontier exhibits clear trade-off:
\begin{itemize}
	\item $\beta_{\text{flow}} \approx 0.50$-$0.75$: Satisfaction approximately 20-30 at txn 50, LRR approximately 1.70-1.76 (customer-focused)
	\item $\beta_{\text{flow}} \approx 1.50$-$2.00$: Satisfaction approximately 0-5 at txn 50, LRR approximately 1.84-1.87 (brand-focused)
	\item $\beta_{\text{flow}} \approx 1.00$-$1.25$: Satisfaction approximately 5-10 at txn 50, LRR approximately 1.80-1.82 (balanced)
\end{itemize}

\begin{figure}[t]
	\centering
	\includegraphics[width=0.5\textwidth]{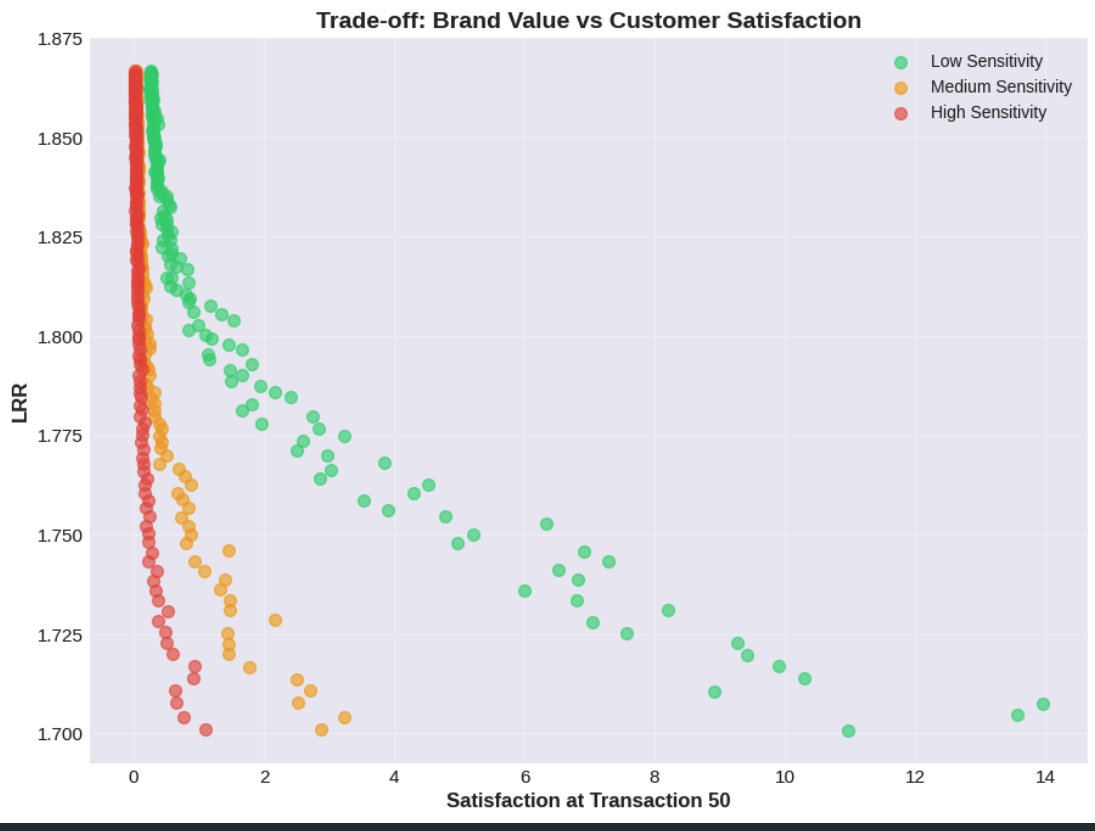}
	\caption{LRR-satisfaction trade-off scatter plot showing Pareto frontier. Each point represents a different $\beta_{\text{flow}}$ value averaged across replications. The frontier demonstrates clear trade-off: higher LRR requires accepting lower satisfaction. No configuration achieves both high LRR and sustained satisfaction in pure outflow scenarios, forcing strategic choice based on brand priorities.}
	\label{fig:exp3_tradeoff_frontier}
\end{figure}

No configuration achieves both high LRR and sustained satisfaction in pure outflow. Strategic profiles must choose their priority:
\begin{itemize}
	\item Conservative profiles ($\beta_{\text{flow}} \approx 0.75$-$1.00$): Accept LRR $\approx 1.76$-$1.80$ to extend satisfaction lifespan by approximately 8-12 transactions through gentler penalty signals
	\item Moderate profiles ($\beta_{\text{flow}} \approx 1.50$): Achieve LRR $\approx 1.84$ while maintaining minimal satisfaction (approximately 1-2) until transaction 20 with moderate penalties
	\item Aggressive profiles ($\beta_{\text{flow}} \approx 2.00$-$2.25$): Maximize LRR $\approx 1.87$-$1.90$ accepting immediate satisfaction collapse through severe penalties
\end{itemize}

\textbf{Critical Implication: Bilateral Flow Necessity.} Experiment 3 demonstrates that pure outflow scenarios (where brands experience only redemption without compensating inflows) are fundamentally unsustainable regardless of parameter configuration. The question is not if satisfaction collapses, but when.

The pricing mechanism does not rebalance pools. It merely penalizes customers for making imbalances worse, hoping to discourage further outflow. The penalties are reactive deterrents, not corrective mechanisms. Actual pool rebalancing requires bilateral flows: one brand's outflow serving as another's inflow. Without compensating inflows from partner brands' customers, penalty pricing can only slow the death spiral, not reverse it.

This finding provides empirical justification for the hybrid architecture's bilateral flow design: sustainable operation requires network-level flow reciprocity, preventing monotonic pool depletion and enabling long-term stability.

In bilateral flow networks, these same parameters exhibit qualitatively different behavior: $\beta_{\text{flow}}$ determines penalty severity that influences flow patterns across the network, rather than futilely attempting to deter inevitable collapse. Understanding how the pricing mechanism performs under realistic bilateral flow conditions represents an essential next step for validating the hybrid architecture's viability.

\subsection{Section Summary: Design Space Foundations}
\label{subsec:section6_summary}

This section established foundational results through systematic analytical exploration:

\textbf{Parameter Independence (Experiment 1).} $\beta_{\text{trans}}$ and $\beta_{\text{flow}}$ exhibit negligible coupling ($<5\%$), enabling independent calibration based on respective design objectives. Brands can separately optimize whale protection and flow rebalancing without concern for interaction effects. This modularity simplifies operational decision-making and enables brands to adjust individual parameters in response to changing conditions.

\textbf{Feasible Design Space (Experiment 2).} At recommended $\theta=0.10$, feasible range $\beta_{\text{flow}} \in [0.69, 2.25]$ balances customer tolerance and protection requirements. Strategic bounds are non-binding for all tested profiles, with threshold choice effectively decoupling bound constraints from parameter selection. This reveals that grace periods provide powerful tools for managing customer experience without sacrificing pool protection capability.

\textbf{Performance Trade-Offs (Experiment 3).} In pure outflow scenarios, $\beta_{\text{flow}}$ controls LRR-satisfaction trade-off but cannot prevent eventual collapse. All configurations yield satisfaction $<50$ within 15-32 transactions, demonstrating that bilateral flow reciprocity constitutes a fundamental architectural requirement rather than an optional enhancement. The pricing mechanism serves as a reactive deterrent, not a corrective rebalancing force.

\textbf{Strategic Profile Differentiation.} Rather than universal optimal values, parameter selection reflects strategic choices balancing customer experience against pool protection. Conservative, moderate, and aggressive profiles represent qualitatively distinct operational philosophies suited to different brand contexts, customer bases, and competitive environments. This design space approach acknowledges the diversity of loyalty program contexts and enables brands to instantiate configurations aligned with their organizational values.

These findings provide clear guidelines for parameter selection while establishing that sustainable operation fundamentally depends on network-level bilateral flows. When one brand's outflow serves as another's inflow, the pricing mechanism can transition from futile resistance against inevitable collapse to effective tool for navigating network equilibrium. Future work validating mechanism performance under realistic bilateral flow conditions will complete the assessment of the hybrid architecture's viability for decentralized multi-brand loyalty networks.

\section{Discussion}
\label{sec:discussion}

Our analysis reveals why coalition loyalty programs consistently fail despite compelling theoretical advantages: the centralized operator model creates structural misalignment where intermediaries extract more value than they provide. This architectural flaw cannot be resolved through better management because the problem lies in how value flows through the system rather than how operations are executed.

The hybrid framework addresses this by eliminating intermediaries entirely. Brands maintain sovereignty while achieving interoperability through protocol-based coordination, transforming what seemed like an inherent trade-off into a solvable design problem. The three-layer pricing mechanism handles value exchange without requiring trusted third parties, demonstrating that sophisticated coordination can emerge from simple, transparent rules.

Three key insights challenge conventional thinking. First, the strategic design space analysis reveals that parameter selection reflects organizational philosophy rather than technical optimization. Conservative and aggressive approaches both work when properly matched to context, acknowledging that loyalty programs span diverse industries with fundamentally different competitive dynamics. Second, the relaxation from eight factors to two broker-observable inputs demonstrates practical implementability - while additional market intelligence would enhance precision when available, the mechanism operates effectively using only trustlessly verifiable data, enabling deployment under real-world constraints rather than requiring ideal conditions. Third, the pure outflow experiments establish that pricing mechanisms influence but cannot create equilibrium, proving bilateral flow reciprocity constitutes an architectural requirement rather than optional enhancement.

This last point extends beyond loyalty programs to mechanism design in decentralized systems generally. Individual components with sophisticated local optimization still face inherent instability when confronting persistent directional pressures. Sustainable coordination requires network-level reciprocity where flows balance naturally, with local mechanisms refining rather than generating equilibrium. Understanding this distinction shapes how we approach any system balancing autonomy with coordination.

Our framework addresses the nine structural challenges dooming traditional coalitions through architectural choices: brands issue their own rewards preserving identity, control partnership networks managing cannibalization, register customers directly maintaining data ownership, operate independently enabling reputation isolation, interact through simple protocols reducing complexity, participate without intermediaries aligning incentives, retain customer relationships enabling personalization, manage their own networks protecting exclusivity, and minimize infrastructure costs ensuring sustainability.

Several limitations warrant acknowledgment. We model steady-state behavior rather than transition dynamics, assume rational actors rather than organizations with political constraints, and examine mechanism design isolated from broader loyalty program functions. Most critically, we analyze pure outflow scenarios providing conservative bounds without validating actual network operation under bilateral flows. These simplifications enabled analytical tractability but mean practical implementation requires further validation addressing how parameter choices perform across interconnected pools, what pool depths prove optimal, and whether brands can operate with less than full backing.

The intelligence-driven commerce transformation makes this work urgent. When AI agents mediate relationships, architectural limitations become absolute barriers. Systems requiring manual coordination or opaque pricing prevent algorithmic optimization in ways protocol-based coordination with transparent pricing do not. As commerce evolves, loyalty programs must correspondingly evolve beyond closed silos and failed coalitions toward architectures recognizing sovereignty and interoperability need not conflict.

\section{Conclusion}
\label{sec:conclusion}

This research makes three primary contributions. First, we demonstrate that coalition loyalty program failures stem from architectural limitations in centralized operator models rather than operational deficiencies, identifying intermediary rent-seeking and misaligned incentives as the root cause. Second, we propose a hybrid framework preserving brand sovereignty while enabling cross-brand utility through trustless protocol-based coordination, addressing structural problems through design rather than management. Third, we derive a complete pricing mechanism integrating market intelligence while enabling fair value exchange without requiring trusted intermediaries.

The framework combines insights across mechanism design, platform economics, and behavioral theory into a coherent system where concerns separate cleanly across layers. Customer prices incorporate dynamic adjustments responding to transaction characteristics and pool conditions. Inter-brand settlement operates through universal asset backing eliminating trust requirements. Compensation flows address externalities ensuring appropriate value attribution. This separation enables independent optimization while maintaining economic coherence.

Most significantly, our pure outflow experiments prove bilateral flow reciprocity constitutes fundamental architectural requirement. Pricing mechanisms influence equilibrium but cannot create it - sustainable operation requires network-level flows where one brand's outflow serves another's inflow. This finding provides empirical justification for the bilateral flow design and establishes direction for future work.

That future work should validate mechanism performance under realistic bilateral flow conditions, examining how parameter choices perform across interconnected pools, what pool depths prove optimal, and whether brands can operate with less than full backing. These questions will determine whether the hybrid architecture transitions from theory to practice, enabling loyalty programs combining the best properties of closed and open systems while avoiding structural problems afflicting both.

The contribution extends beyond loyalty programs to any context balancing participant autonomy with network coordination. When can trustless protocols replace centralized intermediaries? How do we preserve sovereignty while enabling network effects? Our framework provides concrete answers with broad applicability.

The path forward requires moving beyond the false choice between closed silos and failed coalitions. Properly designed decentralized systems can deliver network benefits without centralized control, preserve identity while enabling interoperability, and align incentives through coordination rather than extraction. As commerce transforms toward intelligence-driven paradigms, loyalty programs must evolve correspondingly. The hybrid architecture demonstrates how thoughtful design resolves tensions that seemed inherent while opening possibilities that seemed impossible.

\bibliographystyle{elsarticle-harv} 
\bibliography{references}






\end{document}